\documentclass[11pt]{article}
\usepackage{graphicx}
\usepackage{amssymb}
\usepackage{amscd}
\usepackage{mathrsfs}
\usepackage{longtable,lscape}
\usepackage{amsthm}
\usepackage{amsfonts}
\usepackage{amsmath}
\usepackage{bbm}
\usepackage{float}
\usepackage{url}

\newcommand{\compl}{{\mathbb C}}
\newcommand{\real}{{\mathbb R}}
\newcommand{\captionfonts}{\footnotesize}
\makeatletter  
\long\def\@makecaption#1#2{%
  \vskip\abovecaptionskip
  \sbox\@tempboxa{{\captionfonts #1: #2}}%
  \ifdim \wd\@tempboxa >\hsize
    {\captionfonts #1: #2\par}
  \else
    \hbox to\hsize{\hfil\box\@tempboxa\hfil}%
  \fi
  \vskip\belowcaptionskip}
\makeatother 
\begin{document}
\title{Do Spins Have Directions?}
\author{Diederik Aerts$^1$ and Massimiliano Sassoli de Bianchi$^{2}$ \vspace{0.5 cm} \\ 
        \normalsize\itshape
        $^1$ Center Leo Apostel for Interdisciplinary Studies \\
        \normalsize\itshape
        Brussels Free University, 1050 Brussels, Belgium \\ 
         \normalsize
        E-Mail: \url{diraerts@vub.ac.be}
          \\ 
        \normalsize\itshape
        $^2$ Laboratorio di Autoricerca di Base \\ 
        \normalsize\itshape
         6914 Lugano, Switzerland \\
        \normalsize
        E-Mail: \url{autoricerca@gmail.com} \\
              }
\date{}
\maketitle
\begin{abstract}
\noindent 
The standard Bloch sphere representation has been recently generalized to describe not only systems of arbitrary dimension, but also their measurements, in what has been called the \emph{extended Bloch representation of quantum mechanics}. This model, which offers a solution to the longstanding measurement problem, is based on the \emph{hidden-measurement interpretation of quantum mechanics}, according to which the Born rule results from our lack of knowledge of the measurement interaction that each time is actualized between the measuring apparatus and the measured entity. In this article we present the extended Bloch model and use it to investigate, more specifically, the nature of the quantum spin entities and of their relation to our three-dimensional Euclidean theater. Our analysis shows that spin eigenstates cannot generally be associated with directions in the Euclidean space, but only with generalized directions in the  Blochean space, which apart from the special case of spin one-half entities, is a space of higher dimensionality. Accordingly, spin entities have to be considered as genuine non-spatial entities. We also show, however, that specific vectors can be identified in the Blochean theater that are isomorphic to the Euclidean space directions, and therefore representative of them, and that spin eigenstates always have a predetermined orientation with respect to them. We use the details of our results to put forward a new view of realism, that we call \emph{multiplex realism}, providing a specific framework with which to interpret  the human observations and understanding of the component parts of the world. Elements of reality can be represented in different theaters, one being our customary Euclidean space, and another one the quantum realm, revealed to us through our sophisticated experiments, whose elements of reality, in the quantum jargon, are the eigenvalues and eigenstates. Our understanding of the component parts of the world can then be guided by looking for the possible connections, in the form of partial morphisms, between the different representations, which is precisely what we do in this article with regard to spin entities.
\end{abstract}

\section{Introduction}
\label{Introduction}

According to the hypothesis of realism, reality is out there, available to us to be experienced. Therefore, what we know about reality comes from our experiences of it, and from the worldviews that we are able to construct when we order these experiences into a possibly consistent map of relations. 

No doubts, the human process of creation of a worldview started a long time ago, in pre-cultural and pre-scientific times, when our experiences of reality were only of the ``ordinary kind,'' i.e., obtained using exclusively our human macroscopic body as a measuring apparatus, plus a few very basic tools. From the ordering of these experiences, a first ``clothing and decoration'' of reality resulted, consisting in the identification of those portions of it that were characterizable by aggregates of sufficiently stable properties, like having a certain amount of weight, a given size, shape, state of movement, orientation, etc. These ``clusters of properties'' are what we today call \emph{classical entities}: the macroscopic objects of our everyday life, and the astronomical objects that we can observe moving in the sky, like the Moon and the Sun. 

We can say that this ``clothing and decoration'' of reality happened (and happens) typically in two different ``directions of penetration.'' The first one is a \emph{penetration in depth}, which precisely corresponds to the process we have just mentioned of identifying the entities forming ``stable aggregates of elements of reality.'' The second one is a \emph{penetration in width}, corresponding to a process of organization of the relations existing between the different entities, i.e., between those aggregates of properties that appear to be separate~\cite{Aerts2002} (not influencing each other in a significant way). This penetration in width of our reality can be considered as an ordering process giving rise to a \emph{space}, and more specifically to our \emph{Euclidean three-dimensional space}, which therefore should be considered, first of all, a \emph{space of relations}.

According to the above, it is clear that when a given ensemble of experiences is properly ordered and organized, a specific \emph{theater of reality} will emerge, inside of which a given typology of actors can perform and relate, in predetermined ways. Our Euclidean physical space, in that sense, should not be understood as an external ``container,'' accommodating the play of these classical actors, but really as the manifestation of the structure of their possible relations. It is then also clear that when an ensemble of new experiences becomes all of a sudden available, our first attempt will be that of trying to find a place for them in the theater of reality that we have already constructed. This also because, as time passes by, it is easy to forget about the process of its construction and start believing that all of reality should necessarily fit into it, as the theater and its content, and reality, would just be one and the same thing. 

As is known, an amazing ensemble of new experiences that came shaking the structure of our classical theater are those that were obtained in the investigation of the microscopic layer of our reality, in confirmation of that corpus of knowledge known as \emph{quantum mechanics}. Examples are the famous experiments of \emph{Stern} and \emph{Gerlach} with silver atoms~\cite{Stern1922}, of \emph{Rauch et al.} with neutrons~\cite{Rauch1974, Rauch1975, Rauch1988}, and of \emph{Aspect et al.} with entangled photons~\cite{Aspect1982a, Aspect1999}.  (Another important ensemble of experiments that have undermined the structure of our classical theater are those that have led to the discovery of \emph{relativity theory}, but we shall not be concerned by them in this article).  
Thanks to these very sophisticated experiments, we could access elements of reality that in the past were totally beyond our reach, because of our too coarse physical senses, but also because our natural environment did not provide the controlled conditions of a modern laboratory. These experiments revealed to us that although these new elements of reality were in part similar to those we had so far experienced, at the same time they were sufficiently different to spoil our efforts to incorporate them in our classical \emph{Euclidean theater}.

A typical example is that of spins, which we will discuss more particularly in this article. A spin is usually described as an intrinsic angular momentum carried by a microscopic entity, like an electron, a proton, a neutron, or by a combination of elementary entities (composite entity), like an atomic nuclei. However, although it possesses the same physical dimension of an angular momentum carried by a macroscopic body (the dimension of the Planck constant $\hbar$), it cannot be associated with any specific rotation. For instance, because if the spin is calculated on a classical basis, as a rotation in space, such a rotational movement would yield a superluminal velocity along the microscopic particles' periphery~\cite{Hentschel2009}, in violation of the relativistic limit. Also, in the case of fractionary spins, we know that a $360^{\circ}$ rotation will not bring the spin entity back in the same state~\cite{Rauch1975}, as is instead the case for the actors of the classical theater. 

This impossibility of understanding a spin as a state of rotation in the three-dimensional space can be further evidenced by observing that it cannot generally and consistently be represented as a three-dimensional vector, as is the case for the angular momentum of a macroscopic object. This is so because in quantum mechanics a spin (and more generally an angular momentum) is described by an operator, and therefore cannot be simply drawn as a three-dimensional vector quantity whose components would be real numbers.

To see this more explicitly, consider a general spin-$s$ entity, with $s={1\over 2},1,{3\over 2},2,{5\over 2},\dots$. Its states are described by complex valued vectors in a $N$-dimensional Hilbert space ${\cal H}_N=\compl^N$, with $N=2s+1$. The spin observable $S_{\bf n}$, along the space direction ${\bf n}=n_1\hat{\bf x}_1+n_2\hat{\bf x}_2+n_3\hat{\bf x}_3$, is then given by the product $S_{\bf n} \equiv {\bf S}\cdot {\bf n}=S_1n_1 + S_2n_2 + S_3n_3$, where ${\bf n}$ is a unit vector: $\|{\bf n}\|^2 = n_1^2 + n_2^2 + n_3^2 =1$, and the $\hat{\bf x}_i$, $i=1,2,3$, are the three orthonormal vectors forming the canonical basis of the Euclidean space. The three observables $S_i\equiv {\bf S}\cdot \hat{\bf x}_i$, $i=1,2,3$, are $N\times N$ self-adjoint matrices obeying the commutation relations: $[S_1,S_2]=i\hbar S_3$, $[S_2,S_3]=i\hbar S_1$ and $[S_3,S_1]=i\hbar S_2$, from which one can also deduce that $S^2 \equiv {\bf S}\cdot {\bf S}= S_1^2+S_2^2+S_3^2 = \hbar^2s(s+1) \mathbb{I}$  (this is only true for non-composite spin entities), 
with $\mathbb{I}$ the identity matrix, so that all spin states are also eigenstates of $S^2$, with $N$ times degenerate eigenvalue $\hbar^2s(s+1)$. We also denote by $|\mu,\hat{\bf x}_i\rangle$, $\mu=-s,-s+1,\dots,s-1,s$, the $N$ eignevectors of the spin components $S_i$, $S_i|\mu,\hat{\bf x}_i\rangle=\hbar \mu|\mu,\hat{\bf x}_i\rangle$, which form a basis of ${\cal H}_N$: $\langle \mu,\hat{\bf x}_i|\nu,\hat{\bf x}_i\rangle =\delta_{\mu\nu}$, $\sum_{\mu=-s}^s |\mu,\hat{\bf x}_i\rangle\langle\mu,\hat{\bf x}_i|=\mathbb{I}$, $i=1,2,3$. 

\begin{figure}[!ht]
\centering
\includegraphics[scale =.57]{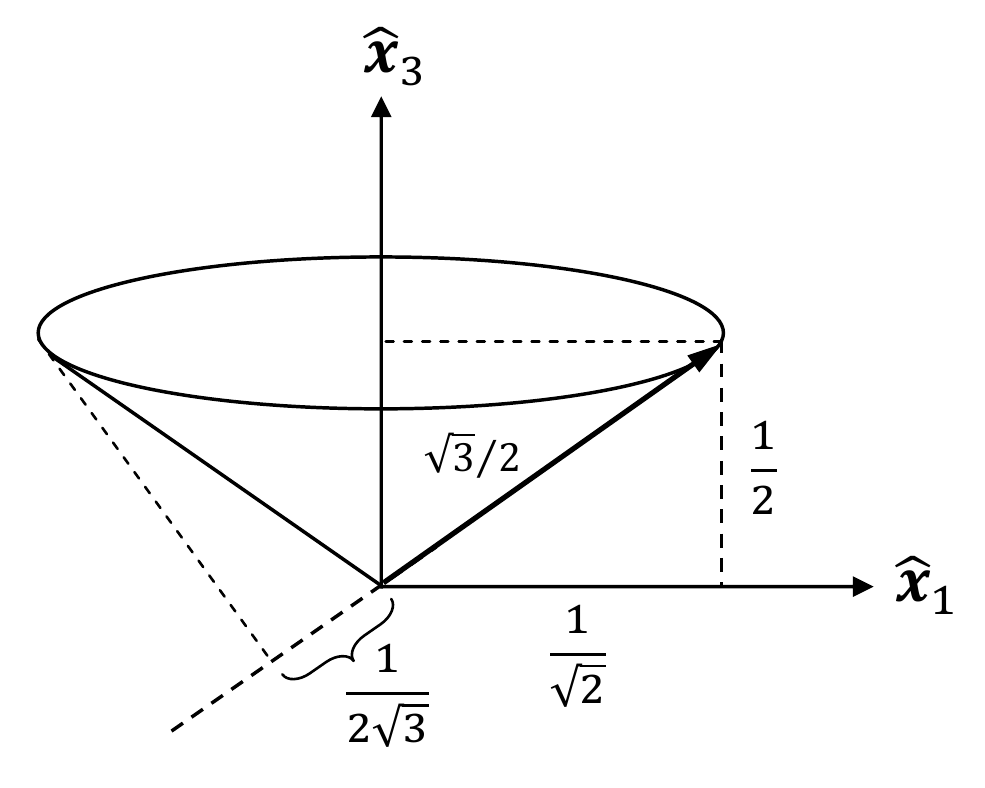}
\caption{The cone representing all the possible classical spin (angular momentum) vectors associable with a spin-${1\over 2}$ entity in the eigenstate $|{1\over 2},\hat{\bf x}_3\rangle$ (in the drawing we have set $\hbar =1$). 
\label{spin3d}}
\end{figure}

When a spin-$s$ entity is, say, in the eigenstate $|\mu,\hat{\bf x}_3\rangle$ of $S_3\equiv {\bf S}\cdot \hat{\bf x}_3$, we can try to classically describe the situation by drawing a vector of length $\hbar \sqrt{s(s+1)}$ (according to $S^2|\mu,\hat{\bf x}_3\rangle= \hbar^2s(s+1)|\mu,\hat{\bf x}_3\rangle$), whose projection onto $\hat{\bf x}_3$ is precisely ${\hbar\mu}$ (according to $S_3|\mu,\hat{\bf x}_3\rangle=\hbar \mu|\mu,\hat{\bf x}_3\rangle$). In fact, there is an entire collection of possible vectors of length $\hbar \sqrt{s(s+1)}$, whose projections onto $\hat{\bf x}_3$ give the value ${\hbar\mu}$: they form a cone of height $\hbar \mu$, lateral height $\hbar \sqrt{s(s+1)}$, and radius $\sqrt{\hbar^2 s(s+1) - \hbar^2 \mu^2}$ (see Fig.~\ref{spin3d}, for the $s={1\over 2}$ case). 

The vectors belonging to the cone, however, cannot consistently represent the spin of a quantum entity in the state $|\mu,\hat{\bf x}_3\rangle$. Indeed, we know that if we measure the spin along any other direction, all the values $-\hbar s,\dots,\hbar s$ are in principle obtainable. This, however, is incompatible with the geometry of the cone. To see this, let us just consider for simplicity the spin-${1\over 2}$ case. If we take one of the cone's vectors as the direction of a spin measurement (here understood in the classical sense), then by orthogonally projecting all the cone's vectors along that specific direction, we obtain the interval of possible values $[-{\hbar\over 2\sqrt{3}}, {\hbar \sqrt{3}\over 2}]$ (see Fig.~\ref{spin3d}), which is manifestly incompatible with the experimental data, as it doesn't contain the eigenvalue $-{\hbar\over 2}$.

The above constitutes a very simple no-go argument showing that it is impossible to directly associate a quantum spin eigenstate (and a fortiori, of course, a superposition of spin eigenstates) with an angluar momentum vector having a specific direction in space,  not even if instead of a single direction we accept to associate to it an entire collection of indeterminate vectors lying on a cone. This means that not only a spin eigenstate cannot be understood as an entity belonging to our three-dimensional spatial theater, it neither can be understood as a higher-dimensional classical-like entity characterized by multiple directions, that would simply ``cast a three-dimensional shadow'' onto it (in a sort of view \`a la \emph{Edwin A. Abbott}).
\begin{figure}[!ht]
\centering
\includegraphics[scale =1.25]{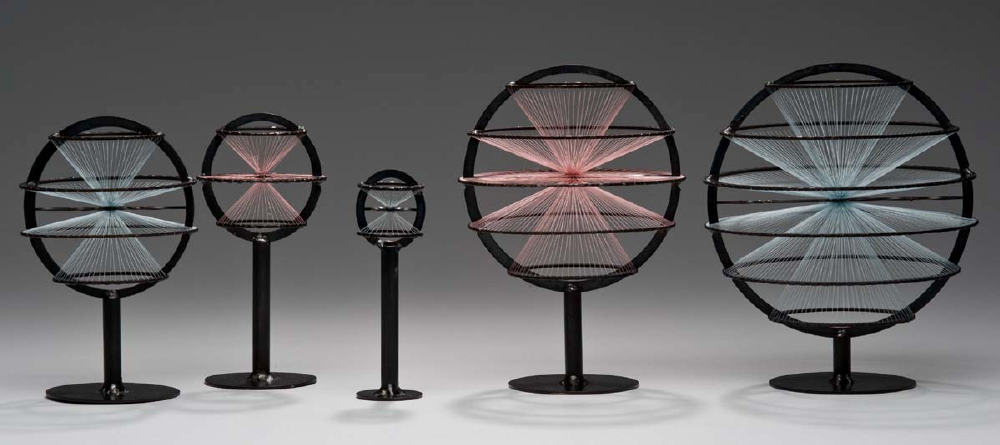}
\caption{Julian Voss-Andreae's \emph{Spin Family} (steel and silk, 2009), is a series of objects displaying the three-dimensional structure of spin entities, with regard to their eigenstates. However, the continuum of directions represented by the cones associated with the different circular metal frames describe ``shadows'' which are not fully compatible with the rules of quantum mechanics, when more than a single measurement is considered.
\label{spinfamily}}
\end{figure} 

So, coming back to our initial discussion, we see that through our quantum experiments we have recently (in the time scale of human evolution) opened a window to an entire new sector of our reality, which requires us to construct a brand new theater, to find a suitable (non-spatial) ``place'' for the new quantum actors, in which their show can be represented. This of course will be quite a non-ordinary theater according to our classical criteria, as these quantum actors can behave in a way that is technically impossible to the classical ones. But considering the existence of these two theaters, the quantum and the classical, two questions  arise. 

The first question is about the nature of their interactions. Indeed, quantum entities, although not in space (and possibly also not in time), are nevertheless available to leave traces into it. If this would not be the case, we would never have learned about their existence, as our measuring apparatuses are macroscopic objects, extensions of our human bodies, which means that we can only explore the content of the quantum theater as from the Euclidean one. In quantum mechanics the effects of these interactions are described by the Born rule and the associated projection postulate, so the question is: \emph{How does a quantum entity interact with a classical one (a measuring apparatus) during a measurement process, so as to produce the different possible outcomes, in accordance with the predictions of the Born rule? Considering that a measurement process cannot be understood as a mere process of discovery of a shadow that a quantum entity would cast onto the Euclidean stage, how can we describe the process through which it leaves a detectable trace onto it?} 

The second question is about finding what are the possible relations between the elements of reality characterizing the entities living in the quantum (non-spatial) theater, and those characterizing the classical entities (spatial objects) located in our human classical theater. In other terms, it is about finding the possible connections between what our sophisticated experimental apparatuses indicate to us to be real, and what we know (or think) to be real from our direct pre-scientific human experience. A more specific formulation of this question, in relation to spin entities, would be the following: \emph{What is the general relation between the non-spatial spin eigenstates and the space directions in the Euclidean space, and is it possible to have a precise geometrical description of such relation?}

The purpose of this article is to offer an answer to the above two questions. For this, the article is organized as follows. In Sec.~\ref{The Bloch sphere}, we introduce the standard Bloch sphere representation, recalling that for spin-${1\over 2}$ entities there is a direct (one-to-one) correspondence between spin states and space directions (each state being oriented along a different space direction). In Sec.~\ref{The extended Bloch sphere}, we show how to extend the Bloch result to also include the measurements, and in Sec.~\ref{General measurement} we explain how the model can be naturally generalized to describe entities of arbitrary dimension $N$, answering in this way the first question. Then, in Sec.~\ref{Spin and space}, we use the extended Bloch representation to show that the spin eigenstates, beyond the spin-${1\over 2}$ case, are not anymore oriented along space directions, and therefore should be considered to be genuine non-spatial elements of our reality (not belonging to our classical theater). However, we also show that the different eigenstates are oriented in a very specific way with respect to the space directions describing the measurements for which they are eigenstates, within the extended Bloch sphere (the quantum theater), answering in this way the second question. In Sec.~\ref{Composition of spins}, we show that this relation between spin eigenstates and space directions remains valid also for composite spin entities, and in Sec.~\ref{Connecting elements of reality} we explain how the correspondence between the elements of reality  in our classical theater, and those in the quantum one, should be understood,  
and what are the limitations with regard to spins. Finally, in Sec.~\ref{Conclusion}, we introduce the notion of \emph{multiplex realism} and offer some concluding remarks.

\section{The Bloch sphere}
\label{The Bloch sphere}

In the previous section we have shown that the representation of  quantum spin eigenstates of different magnitudes by means of cones of vectors in a 3-dimensional space is inconsistent, and therefore misleading, as it conveys the wrong idea that a spin eigenstate could be put into full correspondence with an intrinsic classical angular momentum, although of an indeterminate nature. On the other hand, if we renounce associating spin eigenstates with classical 
angular momenta, it is certainly possible, at least in the special case of spin-${1\over 2}$ entities, to characterize each state by means of a specific direction of space. However, for spin entities of magnitude $s>{1\over 2}$, this will no longer be possible, as to characterize the different eigenstates, a $4s(s+1)$-dimensional quantum theater will be required.

But let us start by considering the simple spin-${1\over 2}$ situation. Let $|\pm{1\over 2},{\bf n}\rangle\in\compl^2$ be the two eigenvectors of the spin operator $S_{\bf n}$, oriented along an arbitrary direction ${\bf n}$, for the eigenvalues $\pm{\hbar\over 2}$. In the following, for simplicity, we will simply write: $|\pm,{\bf n}\rangle\equiv |\pm{1\over 2},{\bf n}\rangle$. Then, also the vector-states $e^{i\alpha}|\pm,{\bf n}\rangle$, with $\alpha\in \real$, are eigenvectors of $S_{\bf n}$, for the same eigenvalues. In other terms, the vectors $|\pm,{\bf n}\rangle$ and $e^{i\alpha}|\pm,{\bf n}\rangle$ represent the same physical states, in accordance with the known fact that no measurement can determine a global phase factor. 

The reason for this is that a measurement is a process that is precisely designed to single out a given entity, in order to specifically measure some of its properties. This means that a measurement is a context conceived to only act on the entity which is meant to be measured, and not, say, on combinations of that entity with other entities. This isolation of the measured entity in the measurement context thus excludes the possibility that a global phase factor would be able to produce interference effects, and therefore become observable.

A way to make fully explicit this isolation of an entity in a measurement context, causing the global phase of its vector-state $|+,{\bf n}\rangle$ to become irrelevant, is to describe its state by means of a one-dimensional projection operator $P({\bf n})=|+,{\bf n}\rangle \langle +,{\bf n}|$. Indeed, knowing $P({\bf n})$, one can always reconstruct $|+,{\bf n}\rangle$, but only up to a global phase factor. To show this, we write $|+,{\bf n}\rangle = a_+|+,\hat{\bf x}_3\rangle + a_-|-,\hat{\bf x}_3\rangle$, so that (to simplify the notation, we set: $|\pm\rangle\equiv |\pm,\hat{\bf x}_3\rangle$): $P({\bf n})= |a_+|^2 |+\rangle\langle +| + |a_-|^2 |-\rangle\langle -|+ (a_+a_-^*|+\rangle\langle -| + {\rm c.c.})$. We assume that $a_+\neq 0$, and we fix the global phase of $|+,{\bf n}\rangle$ in a way that $a_+$ is a real strictly positive number. Then, since $\langle +|P({\bf n})|+\rangle= |a_+|^2 = a_+^2$, we have $a_+=\langle +|P({\bf n})|+\rangle^{1\over 2}$. Also, $\langle +|P({\bf n})|-\rangle=a_+a_-^*$, so that $a_-=\langle -|P({\bf n})|+\rangle \langle +|P({\bf n})|+\rangle^{-{1\over 2}}$. And of course, if $a_+=0$, then $a_-$ is simply a phase factor.

So, when we go from the \emph{vector-state} representation $|+,{\bf n}\rangle$, to the associated (rank-1 projection) \emph{operator-state} representation $P({\bf n})$, the only information we loose is that relative to the vector's global phase, which plays no role in a measurement. Therefore, the operator-state $P({\bf n})$ can be considered to be the proper mathematical object representing the state in which the spin entity (and more generally any quantum entity) is prepared and isolated in view of being subjected to a measurement process. 

Our next step is to show how to represent the spin states $P({\bf n})$ as directions in space. This is a well-known representation that dates back to 1892, when the French physicist and mathematician \emph{Henri Poincar\'e} discovered that a surprisingly simple representation of the polarization states of electromagnetic radiation could be obtained by representing the polarization ellipse on a complex plane, and then further projecting such plane onto a sphere~\cite{Poincare1892}. In 1946, this representation was adapted by the Swiss physicist \emph{Felix Bloch}~\cite{Bloch1946} to represent the states of two-level quantum systems, like spin-${1\over 2}$ entities, in what is today known as the \emph{Poincar\'e sphere} or \emph{Bloch sphere}. 

This representation is a consequence of the observation that a general complex $2\times 2$ matrix can always be written as a linear combination of four matrices $\sigma_i$, $i=0,1,2,3$, where $\sigma_0 =\mathbb{I}$, is the identity matrix, and $\sigma_i={2\over\hbar}S_j$, $j=1,2,3$, are the three Pauli matrices. In the eigenbasis of $S_3$, these four matrices take the explicit form: 
\begin{equation}
\sigma_0=
\left[ \begin{array}{cc}
1 & 0 \\
0 & 1 \end{array} \right],\quad
\sigma_1=
\left[ \begin{array}{cc}
0 & 1 \\
1 & 0 \end{array} \right],\quad
\sigma_2=
\left[\begin{array}{cc}
0 & -i \\
i & 0 \end{array} \right],\quad 
\sigma_3=
\left[ \begin{array}{cc}
1 & 0 \\
0 & -1 \end{array} \right].
\label{Pauli-matrices}
\end{equation}
It is then straightforward to show that they are all mutually orthogonal, in the sense that ${\rm Tr}\,\sigma_i\sigma_j = 2\delta_{ij}$, from which it follows that (setting $i=0$): ${\rm Tr}\,\sigma_j=0$, $j=1,2,3$. 

Therefore, the state $P({\bf n})$ can generally be written as the linear combination of the four matrices $\sigma_i$: $P({\bf n})={1\over 2}(a_0\sigma_0 + a_1\sigma_1 + a_2\sigma_2+ a_3\sigma_3)$, where the factor ${1\over 2}$ has been introduced for convenience. Since $P({\bf n})$ and the $\sigma_i$ are all self-adjoint, the coefficients $a_i$ must be real. To determine them, we observe that $a_i={\rm Tr}\,P({\bf n})\sigma_i$, from which it follows that $a_0=1$, since ${\rm Tr}\,P({\bf n})=1$. Considering that $P({\bf n})^2=P({\bf n})$, we have ${\rm Tr}\,P^2({\bf n})=1$, implying that $a_1^2+a_2^2+a_3^2=1$. We also know that $P({\bf n})$ obeys the eigenvalue equation $S_{\bf n}P({\bf n})={\hbar\over 2}P({\bf n})$, which we can write as $(n_1\sigma_1 + n_2\sigma_2 + n_3\sigma_3)P({\bf n}) = P({\bf n})$. Taking the trace of this expression, we thus find that: $n_1a_1 + n_2a_2 + n_3a_3 = 1$, from which we conclude that $a_i=n_i$, $i=1,2,3$. This means that $P({\bf n})$ is fully determined by the unit real vector ${\bf n}$, and we can write in compact form:
\begin{equation}
P({\bf n})={1\over 2}(\mathbb{I} + {\bf n}\cdot \mbox{\boldmath$\sigma$}),
\label{2-d-formula}
\end{equation}
where $\mbox{\boldmath$\sigma$}\equiv(\sigma_1,\sigma_2,\sigma_3)^T$. In particular, if we write the unit vector ${\bf n}$ in spherical coordinates, that is: ${\bf n}=(\sin\theta\cos\phi, \sin\theta\sin\phi, \cos\theta)^T$, the operator-state $P({\bf n})\equiv P(\theta,\phi)$ takes the explicit form: 
\begin{equation}
P(\theta,\phi)=
{1\over 2}\left[ \begin{array}{cc}
1 +\cos\theta & \sin\theta (\cos\phi -i \sin\phi) \\
\sin\theta (\cos\phi +i \sin\phi) & 1-\cos\theta \end{array} \right] = 
\left[ \begin{array}{cc}
\cos^2{\theta\over 2} & \sin{\theta\over 2} \cos{\theta\over 2}e^{-i\phi} \\
\sin{\theta\over 2} \cos{\theta\over 2} e^{i\phi} & \sin^2{\theta\over 2} \end{array} \right].
\label{spherical-coordinates}
\end{equation}

\begin{figure}[!ht]
\centering
\includegraphics[scale =.57]{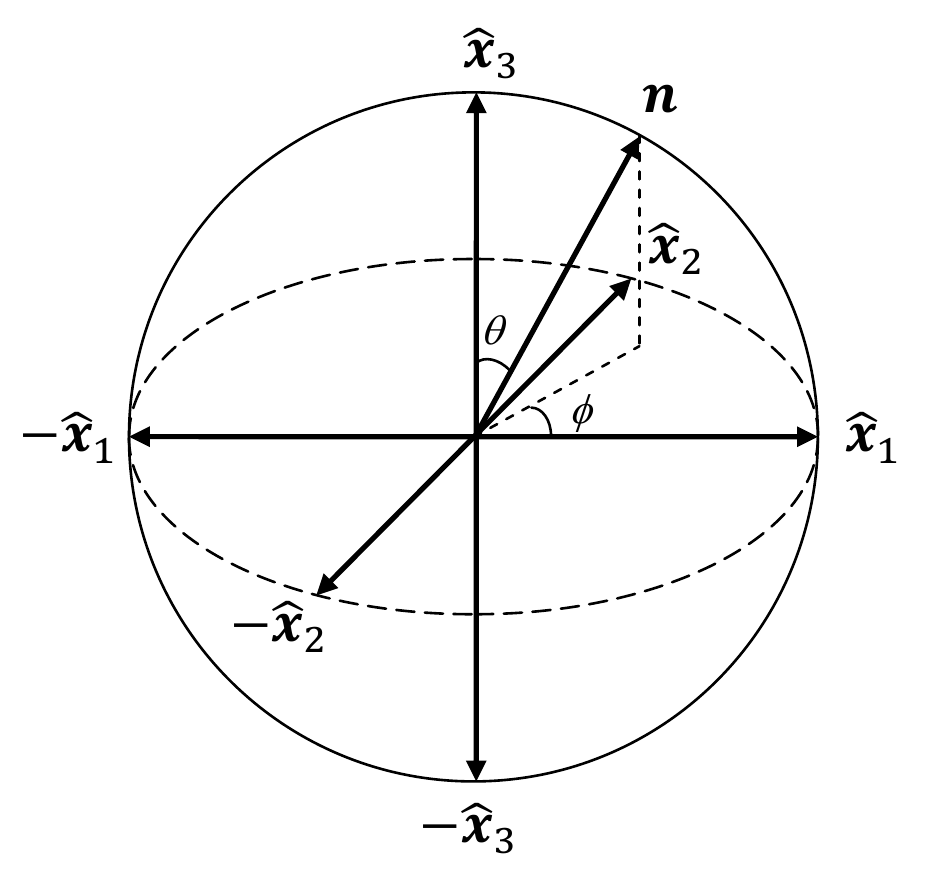}
\caption{The representation of the state of a two-dimensional quantum entity, as a real vector ${\bf n}$ belonging to the surface of a unit sphere (called the Bloch sphere). More precisely, the two opposed unit vectors $\pm{\bf n}$ represent the operator-states $P(\pm{\bf n})={1\over 2}(\mathbb{I} \pm {\bf n}\cdot \mbox{\boldmath$\sigma$})$, which are the two eigenstates of the spin observable $S_{\bf n}$, for the eigenvalues $\pm{\hbar\over 2}$.
\label{Bloch-sphere}}
\end{figure}

The spin eigenstates of a spin-${1\over 2}$ entity can thus be fully determined by specifying the unit vectors ${\bf n}$ defining the space direction for which the associated spin observable $S_{\bf n}$, if measured, would produce with certainty the outcome ${\hbar\over 2}$. This representation (see Fig.~\ref{Bloch-sphere}) is however very different from the (incomplete) cone representation described in the previous section, as is clear that none of the classical spin vectors lying on the cone point to the same direction as the unit vector representative of the state. 
Apart from that difference, the Bloch sphere allows us to maintain a one-to-one correspondence between spin states and space directions, and is valid for arbitrary measurements. However, such direct correspondence will be lost when considering spin entities of magnitude greater than ${1\over 2}$, in the sense that the directions of space will not anymore be sufficient to characterize all the possible relations between states and measurements, which can only be represented in an extended space of higher dimensionality. However, each spin eigenstate will maintain a very specific relation (orientation) with respect to the space direction characterizing its measurement. But before showing this, we need to extend the Bloch representation.

\section{The extended Bloch sphere}
\label{The extended Bloch sphere}

In the previous section we have described the well-known Bloch sphere representation for spin-${1\over 2}$ entities (also valid for general qubit systems). This is a representation that works for the states, and not for the measurements. However, it is possible to extend it to also include a modelization of the different possible measurements, as was shown by one of us forty years after the work of Bloch, providing at the same time a plausible explanation of the origin of the Born rule~\cite{Aerts1986,Aerts1987}. 

Before showing in this section how this can be done, it is worth mentioning that the perspective we are taking here is that which consists in taking the quantum formalism very seriously. By this we mean that we consider that a Hilbertian vector-state, or (when its global phase becomes irrelevant) its associated operator-state, does actually describe the real state of the physical entity under investigation, and not just our knowledge or our beliefs about its condition. Accordingly, we also consider that a measurement process is an objective physical process, bringing the entity from an initial pre-measurement state to a final post-measurement state, in a way that cannot be determined in advance. However, what can be determined in advance are the probabilities of the different transitions, producing the different possible outcomes of the measurement.

Once the quantum formalism is taken seriously in the above sense, we have of course to face the challenge of understanding what happens during the quantum measurement process. Indeed, as is well known, measurements are not explained in the standard quantum formalism, as the Born rule of correspondence, with which the different probabilities are calculated, is just postulated. 

To be more specific, consider a spin-${1\over 2}$ entity prepared in state $P({\bf n})$, subjected to the measurement of the observable $S_3$. According to the projection postulate, two transitions can possibly occur during the measurement process: the transition $P({\bf n})\to P(\hat{\bf x}_3)$, associated with the eigenvalue ${\hbar\over 2}$, or the transition $P({\bf n})\to P(-\hat{\bf x}_3)$, associated with the eigenvalue $-{\hbar\over 2}$. The Born rule then tells us that the probabilities associated with these two transitions are: 
\begin{eqnarray}
{\cal P}(P({\bf n})\to P(\pm\hat{\bf x}_3)) &=& {\rm Tr}\, P({\bf n})P(\pm\hat{\bf x}_3) = {1\over 4}{\rm Tr}\,(\mathbb{I} + {\bf n}\cdot \mbox{\boldmath$\sigma$})(\mathbb{I} \pm\hat{\bf x}_3\cdot \mbox{\boldmath$\sigma$})\\
&=&{1\over 2} \pm {1\over 4}{\rm Tr}\, ({\bf n}\cdot \mbox{\boldmath$\sigma$})(\hat{\bf x}_3\cdot \mbox{\boldmath$\sigma$})={1\over 2}(1\pm {\bf n}\cdot \hat{\bf x}_3),
\label{transition-probabilities}
\end{eqnarray}
where for the third equality we have used the fact that the Pauli matrices have zero trace, and for the last one that they are mutually orthogonal. It is possible to write the above expressions in a more compact form by introducing the polar angle $\theta$ between ${\bf n}$ and $\hat{\bf x}_3$, given by $\cos\theta={\bf n}\cdot \hat{\bf x}_3$ (see Fig.~\ref{Bloch-sphere}). This gives: 
\begin{equation}
{\cal P}(P({\bf n})\to P(\hat{\bf x}_3)) = \cos^2 {\theta\over 2},\quad {\cal P}(P({\bf n})\to P(-\hat{\bf x}_3))= \sin^2 {\theta\over 2}.
\label{transition-probabilities2}
\end{equation}

If the states $P({\bf n})$ are assumed to express an objective condition of the physical entity, it is clear that a measurement has also to be understood not just as a process of \emph{discovery}, but also as a process of \emph{creation}. The discovery aspect is related to the fact that the statistics of outcomes depends on the pre-measurement state (which for instance can be recovered by quantum tomography techniques); the creation aspect, on the other hand, is related to the fact that the process literally creates the property that is observed, in a non-predictable way (apart of course in the special situation where the pre-measurement state is an eigenstate of the measurement). 

What we will now show is how the measurement of an observable, for example the spin observable $S_3$, producing the probabilities (\ref{transition-probabilities2}), can be modeled within the Bloch sphere, in what is called the \emph{extended Bloch representation}~\cite{AertsSassoli2014c}. To do so, we will need more (pure) states than those that are usually considered in the standard formalism. Indeed, to describe the indeterministic dynamics characterizing a quantum measurement, and derive from it the Born rule, we need to go inside the Bloch sphere, i.e., to also consider non-unit representative vectors.

We start by specifying to what kind of operators the non-unit vectors ${\bf r}$, living inside the sphere, are associated with. Let $D({\bf r})={1\over 2}(\mathbb{I} + {\bf r}\cdot \mbox{\boldmath$\sigma$})$. Clearly, even if $\|{\bf r}\|< 1$, we still have ${\rm Tr}\, D({\bf r})=1$. Also, considering that the Pauli matrices have eigenvalues $\pm 1$, it follows that $D({\bf r})$ has eigenvalues ${1\over 2}(1\pm\|{\bf r}\|)$, and therefore is a positive semidefinite operator. Being also a manifestly self-adjoint operator (${\bf r}$ is a real vector), it corresponds to what is usually called a \emph{density matrix} and is generally interpreted as a classical statistical mixture of pure states. However, such interpretation is not without difficulties, as a same density matrix can have arbitrarily many different representations as a mixture of one-dimensional projection operators~\cite{Hughston1993,AertsSassoli2014c}. Our assumption here is that density matrices are able to describe not only classical mixtures of states, but also (extended) pure states, and precisely those pure states that describe how an entity, during a measurement process, approaches the measurement's ``region of potentiality,'' triggering in this way the actualization of a specific interaction. 

\begin{figure}[!ht]
\centering
\includegraphics[scale =.57]{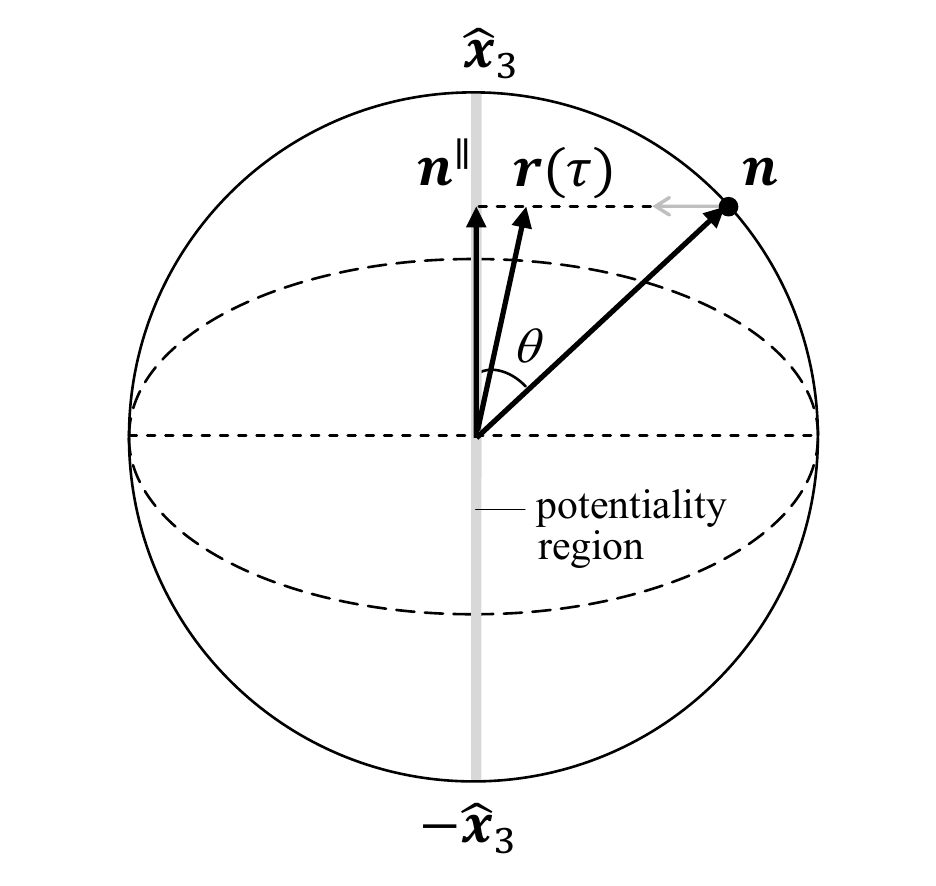}
\caption{The orthogonal path ${\bf r}(\tau)$ followed by the point particle representative of the measured entity in the Bloch sphere, going from the initial position ${\bf r}(0)={\bf n}$, at its surface, to the position ${\bf r}(1)={\bf n}^\parallel$, on the ``potentiality region'' (the gray line segment), here describing the measurement of $S_3$.
\label{Bloch-sphere2}}
\end{figure}

Since the Bloch sphere is a sphere of states, to geometrically represent measurements inside of it we have to consider their associated eigenstates. In the case of the observable $S_3$, we have the two eigenstates $P(\hat{\bf x}_3)$ and $P(-\hat{\bf x}_3)$, represented in the sphere by the two zenith and nadir points $\hat{\bf x}_3$ and $-\hat{\bf x}_3$, respectively. In addition to these two antipodal points, we consider all the intermediary points forming the one-dimensional segment of length 2 going from $-\hat{\bf x}_3$ to $\hat{\bf x}_3$, corresponding to the vertical diameter of the sphere (see Fig.~\ref{Bloch-sphere2}). 

This line segment, or diameter, associated with $S_3$, precisely corresponds to that \emph{potentiality region} characterizing the measurement context, which is responsible for the indeterministic ``collapse'' (as we shall see in a moment). But in order for this to occur, the entity has first to enter into contact with that region. This means that the measurement process has to involve a preparatory \emph{deterministic} phase, through which the pre-measurement state $P({\bf n})$ is brought into contact with the latter. This process corresponds to the immersion of a point particle associated with the vector ${\bf n}$, representative of the entity's pre-measurement state, from the surface of the sphere to its interior, to reach the line segment representing the potentiality region, along an orthogonal path (see Fig.~\ref{Bloch-sphere2}).
If we denote ${\bf n}^\parallel$ the point on the line segment obtained in this way, that is, ${\bf n}^\parallel=({\bf n}\cdot \hat{\bf x}_3)\,\hat{\bf x}_3 = \cos \theta\, \hat{\bf x}_3$, we can describe this deterministic movement of approach of the potentiality region by means of a parameter $\tau$, which is varied from 0 to 1: ${\bf r}(\tau) = (1-\tau)\,{\bf n} + \tau\, \cos \theta\, \hat{\bf x}_3$. Clearly, ${\bf r}(0) = {\bf n}$ is the initial condition, and ${\bf r}(1)={\bf n}^\parallel=\cos \theta\, \hat{\bf x}_3$ is the final condition on the potentiality region. 

Before describing the purely indeterministic part of the measurement, it is instructive to write the generalized operator-state $D({\bf r}(\tau))$ in explicit terms. With ${\bf n}=(\sin\theta\cos\phi, \sin\theta\sin\phi, \cos\theta)^T$ and $\hat{\bf x}_3=(0, 0, 1)^T$, we have ${\bf r}(\tau)=((1-\tau)\sin\theta\cos\phi, (1-\tau)\sin\theta\sin\phi, \cos\theta)^T$, so that $D({\bf r}(\tau))\equiv D(\tau,\theta,\phi)$ takes the explicit form: 
\begin{equation}
D(\tau,\theta,\phi)=
\left[ \begin{array}{cc}
\cos^2{\theta\over 2} & (1-\tau)\sin{\theta\over 2} \cos{\theta\over 2}e^{-i\phi} \\
(1-\tau)\sin{\theta\over 2} \cos{\theta\over 2} e^{i\phi} & \sin^2{\theta\over 2} \end{array} \right].
\label{decoherence}
\end{equation}
As it can be seen in this explicit expression, the deterministic approach of the entity's state towards the potentiality region corresponds to a decoherence-like process, causing the off-diagonal elements of the operator-state $D(\tau,\theta,\phi)$ to gradually vanish, as $\tau\to 1$, so that the ``on-potentiality region'' state has the form of a fully reduced density operator: $D({\bf n}^\parallel) = D(1,\theta,\phi) = \cos^2{\theta\over 2}\, P(\hat{\bf x}_3) + \sin^2{\theta\over 2}\, P(-\hat{\bf x}_3)$.

\begin{figure}[!ht]
\centering
\includegraphics[scale =.57]{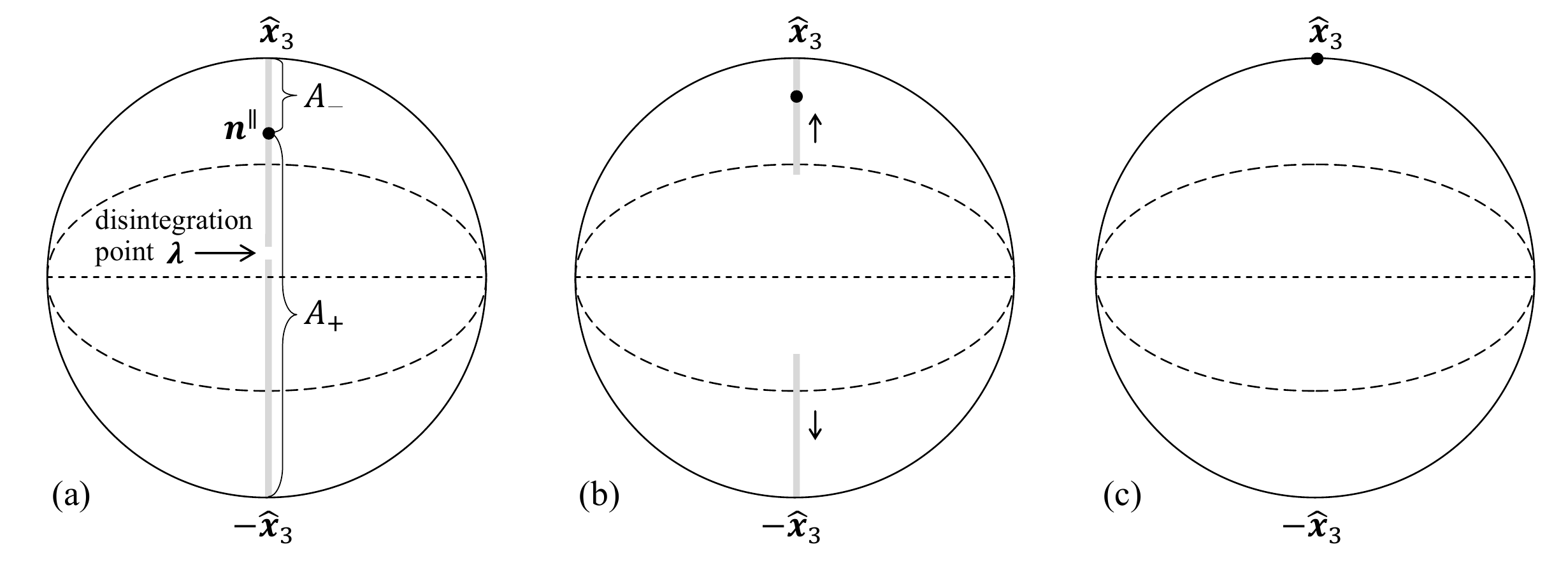}
\caption{The unfolding of the indeterministic part of the measurement of the observable $S_3$: (a) the point particle representative of the state of the spin entity reaches the potentiality region (the one-dimensional elastic substance represented here in gray color) at point ${\bf n}^\parallel$, so defining two regions $A_-$ and $A_+$. The elastic substance then disintegrate at some unpredictable point \mbox{\boldmath$\lambda$}, here assumed to be within region $A_+$; (b) the elastic substance collapses, drawing the point particle towards one of the two anchor points, here $\hat{\bf x}_3$; (c) the point particle reaches its final destination, here point $\hat{\bf x}_3$, representative of the eigenstate $P(\hat{\bf x}_3)$, associated with the eigenvalue ${\hbar\over 2}$.
\label{Bloch-sphere3}}
\end{figure}

We now describe the second phase of the measurement process, corresponding to that purely indeterministic dynamics which is responsible for the emergence of the quantum probabilities, in accordance with the Born rule. For this, we have to think of the potentiality region as a region made of a uniform substance which is not only \emph{attractive} (as it causes the initial state to be orthogonally attracted towards it), but also \emph{unstable} and \emph{elastic}. A simple image we can use is that of a uniform elastic band that would have been stretched between the two anchor points $-\hat{\bf x}_3$ and $\hat{\bf x}_3$, with the state of the entity represented by a point particle stuck onto it, at point ${\bf n}^\parallel$. The instability of the substance means that the elastic at some moment will break, at some unpredictable point \mbox{\boldmath$\lambda$}, so causing its splitting into two halves, which will then contract towards the respective anchor points. Depending on which of these two halves the point particle is attached to, it will either be drawn to point $-\hat{\bf x}_3$, or to point $\hat{\bf x}_3$ (see Fig.~\ref{Bloch-sphere3}).

More precisely, let $A_+$ be the line segment between ${\bf n}^\parallel$ and $-\hat{\bf x}_3$, and $A_-$ be the line segment between $\hat{\bf x}_3$ and ${\bf n}^\parallel$. Their lengths (Lebesgue measures) are: 
\begin{eqnarray}
&&\mu(A_+)= \|{\bf n}^\parallel - (-\hat{\bf x}_3)\|= \|\cos \theta\, \hat{\bf x}_3 + \hat{\bf x}_3\| = \|(1+ \cos \theta)\,\hat{\bf x}_3\| =1+ \cos \theta,\\
&&\mu(A_-)= \|\hat{\bf x}_3 - {\bf n}^\parallel \|=\| \hat{\bf x}_3 - \cos \theta\, \hat{\bf x}_3 \| = \|(1- \cos \theta)\,\hat{\bf x}_3\| =1- \cos \theta.
\end{eqnarray}
Considering that the first immersive phase of the measurement process is perfectly deterministic, it is clear that the transition probability ${\cal P}(P({\bf n})\to P(\pm\hat{\bf x}_3))$ is nothing but the probability that the point particle is drawn to point $\pm\hat{\bf x}_3$, which in turn corresponds to the probability ${\cal P}(\mbox{\boldmath$\lambda$}\in A_\pm)$ that the disintegration point \mbox{\boldmath$\lambda$} belongs to region $A_\pm$. Being the elastic substance, by hypothesis, uniform, and of total length $\mu(A_+)+\mu(A_-)=2$, we thus have: 
\begin{eqnarray}
&&{\cal P}(\mbox{\boldmath$\lambda$}\in A_+) = {1\over 2}\,\mu (A_+) = {1\over 2}(1+ \cos \theta) = \cos^2{\theta\over 2},\\
&&{\cal P}(\mbox{\boldmath$\lambda$}\in A_-) = {1\over 2}\,\mu (A_-) = {1\over 2}(1- \cos \theta) = \sin^2{\theta\over 2},
\label{PlambdainA+}
\end{eqnarray}
which are precisely the quantum mechanical probabilities (\ref{transition-probabilities2}) predicted by the Born rule. In other terms, the standard Bloch sphere representation can be extended to also include a description of the different possible measurements, in accordance with the predictions of the Born rule. 

Before explaining in the next section how this representation can be further generalized to an arbitrary number of dimensions, a few remarks are in order. What we have described is clearly a measurement of the first kind. Indeed, once the point particle has reached one of the two outcome positions $\pm\hat{\bf x}_3$, if subjected again to the same measurement, being already located in one of the two anchor points of the elastic, we have ${\bf n} = {\bf n}^\parallel = \pm\hat{\bf x}_3$, so that its position cannot be further changed by its collapse. 

The disintegration points \mbox{\boldmath$\lambda$} can be interpreted as variables specifying the \emph{measurement interactions}. Thus, the model provides a consistent hidden-measurement interpretation of the quantum probabilities, as epistemic quantities characterizing our lack of knowledge regarding the measurement interaction that is actualized between the measured entity and the measuring apparatus, at each run of the experiment.

It is worth observing that almost each measurement interaction \mbox{\boldmath$\lambda$} gives rise to a purely deterministic process, changing the state of the entity from ${\bf n}^\parallel$ to either $\hat{\bf x}_3$ or $-\hat{\bf x}_3$, depending whether $\mbox{\boldmath$\lambda$}\in A_+$, or $\mbox{\boldmath$\lambda$}\in A_-$. We say `almost' because when $\mbox{\boldmath$\lambda$}={\bf n}^\parallel$, that is, when \mbox{\boldmath$\lambda$} coincides with the point separating region $A_-$ from region $A_+$, we have a situation of classical unstable equilibrium. This point of classical instability is at the origin of the distinction between the two outcomes, but being of measure zero, it doesn't contribute to the values of the probabilities associated with them. In other terms, although the border points $\mbox{\boldmath$\lambda$}={\bf n}^\parallel$ are the ``source of the possibilities,'' they do not contribute to the values of the probabilities that are associated with them.

\section{Modeling a general measurement}
\label{General measurement}

The preliminary 1986 study that allowed to extend the 3-dimensional Bloch sphere representation to also include measurements (as shown in the previous section) generated over the years a number of works, further exploring the explicative power contained in this modelization, also called the \emph{hidden-measurement interpretation} of quantum mechanics (see~\cite{AertsSassoli2014c} and the references cited therein). According to it, a quantum measurement is an experimental context characterized by a lack of knowledge not about the state of the measured entity, but about the interaction between the measuring apparatus and the measured entity. And it is precisely through a mechanism of \emph{actualization of potential interactions} (or of spatialization of non-spatial interactions) that the non-ordinary entities belonging to the quantum theater would leave their ephemeral traces into our classical spatial theater (the first question we addressed in the Introduction). 

For some time this hidden-measurement extension of the Bloch sphere, called the \emph{$\epsilon$-model}~\cite{{Aerts1998b,Aerts1999,AertsSassoli2014c}} (which also allowed an exploration of the intermediary region between classical and quantum) was taken in serious consideration only by a small number of physicists working in the foundations of physical theories. This probably because of the existence of the so-called no-go theorems, like those of Gleason~\cite{Gleason1957} and Kochen-Specker~\cite{Kochen1967}, which were known to be valid only for $N$-dimensional Hilbert spaces with $N>2$, hence not for the special case of two-level systems (qubit), which therefore were considered to be pathological. Accordingly, the extended Bloch representation was mostly considered as a mathematical curiosity, but this was a misconception, as the hidden-measurement interpretation has little to do with a classical hidden-variable theory, as is clear that the lack of knowledge is not associated with the state of the measured entity, but with its measurement-interactions (so that the no-go theorems do not apply). 

Despite this prejudice that no hidden-measurement modelizations could be given for dimensions $N>2$, new results became available over the years, showing that the hidden-measurement mechanism was by no means restricted to two-dimensional situations~\cite{Aertsetal1997, Coecke1995, Coecke1995b}. These promising results were however not totally convincing, as what was still lacking was a natural generalization of the Bloch sphere representation beyond the two-dimensional situation. Things changed in more recent times, when some important mathematical results became available, precisely providing this much-sought generalized Bloch sphere representation, which exploits the properties of the so-called generators of $SU(N)$, the special unitary group of degree $N$~\cite{Arvind1997, Kimura2003, Byrd2003, Kimura2005, Bengtsson2006, Bengtsson2013}. 

Thanks to these results, very recently we could extend the generalized Bloch construction to also include a full description of the measurements (including the degenerate ones), offering in this way what we think is a general and convincing solution to the measurement problem~\cite{AertsSassoli2014c}. In this section we will explain how a $N$-outcome measurement can precisely be described, and therefore explained, in this extended Bloch modelization. In other terms, we will now consider an entity whose Hilbert state is ${\cal H}_N=\compl^N$ (if it is a spin-$s$ entity, then $s={N-1\over 2}$). 

What we need to observe is that, as already mentioned, the 3-dimensional Bloch-sphere representation is based on the property of the Pauli matrices and the identity operator of generating a basis for all linear operators on ${\mathbb C}^2$, so allowing to expand any operator-state on them, in the form: $D({\bf r}) = {1\over 2}\left(\mathbb{I} + {\bf r}\cdot\mbox{\boldmath$\sigma$}\right)$. This correspondence between $2\times 2$ operator-states and real vectors in the three-dimensional unit ball is the expression of the well-known homomorphism between $SU(2)$ and $SO(3)$. Such homomorphism cannot be extended beyond the $N=2$ situation, but one can still represent operator-states as real vectors in a unit sphere, which however will not be anymore three-dimensional, nor completely filled with states. 

More precisely, the reason why operator-states can still be represented in this way is that one can always find $N^2-1$ self-adjoint $N\times N$ matrices $\Lambda_i$, $i=1,\dots,N^2-1$, of zero trace, which together with the identity operator form a basis for all linear operators acting on ${\mathbb C}^N$. These matrices are the so-called generators of $SU(N)$, and if we choose the normalization ${\rm Tr}\,\Lambda_i\Lambda_j=2\delta_{ij}$, a convenient construction is~\cite{Hioe1981, Alicki1987, Mahler1995}: $\{\Lambda_i\}_{i=1}^{N^2-1}=\{U_{jk},V_{jk},W_{l}\}$, with:
\begin{eqnarray}
\label{rgeneratorsN}
&&U_{jk}=|b_j\rangle\langle b_k| + |b_k\rangle\langle b_j|, \quad V_{jk}=-i(|b_j\rangle\langle b_k| - |b_k\rangle\langle b_j|),\\
&&W_l =\sqrt{2\over l(l+1)}\left(\sum_{j=1}^l |b_j\rangle\langle b_j|-l|b_{l+1}\rangle\langle b_{l+1}|\right),\label{generatorsW}\label{W}\\
&&1\leq j < k\leq N, \quad 1\leq l\leq N-1,
\end{eqnarray}
where $\{|b_1\rangle,\dots,|b_N\rangle\}$ is an arbitrary orthonormal basis of $\compl^N$. 

For the two-dimensional case ($N=2$), the generators are $2^2-1=3$, and if $\{|b_1\rangle,|b_2\rangle\}$ is taken to be the canonical basis, they precisely correspond to the Pauli matrices (\ref{Pauli-matrices}). For the three-dimensional case ($N=3$), the generators are $3^2-1=8$, and correspond to the so-called Gell-Mann matrices. So, a general operator-state can always be written as the linear combination:
\begin{equation}
D({\bf r}) = {1\over N}\left(\mathbb{I} +c_N\, {\bf r}\cdot\mbox{\boldmath$\Lambda$}\right) = {1\over N}\left(\mathbb{I} + c_N\sum_{i=1}^{N^2-1} r_i \Lambda_i\right),
\label{formulaNxN}
\end{equation}
where for convenience we have introduced the constant $c_N\equiv \sqrt{N(N-1)\over 2}$. Taking the trace of $D^2({\bf r}) $, then using the orthogonality of the $\Lambda_i$, after a simple calculation one finds that: ${\rm Tr}\, D^2({\bf r})={1\over N} + (1-{1\over N})\|{\bf r}\|^2$, and since an operator-state generally obeys ${1\over N}\leq {\rm Tr}\, D^2({\bf r})\leq 1$, it follows that $0\leq \|{\bf r}\|^2\leq 1$, that is, that the representative vectors ${\bf r}$ belong to a $(N^2-1)$-dimensional unit ball $B_1(\real^{N^2-1})$. Clearly, $\|{\bf r}\|=1$ iff $D^2({\bf r})=D({\bf r})$, i.e., if $D({\bf r})$ is a one-dimensional projection operator (a pure state, in the standard quantum terminology). 

Similarly to the $N=2$ situation, states can therefore be represented as vectors living in a unit ball. On its surface, one finds the rank-one projection operators, and in its interior the density-matrices, i.e., the operator-states. The important difference with the standard Bloch representation is that the ball is now $(N^2-1)$-dimensional, i.e., it is not anymore representable in $\real^{3}$, and, more importantly, only a small portion of it contains states. Indeed, if it is true that for every operator-state $D$ one can always find a vector ${\bf r}\in B_1(\real^{N^2-1})$ such that $D$ can be written in the form (\ref{formulaNxN}), the converse will not be generally true: given a vector ${\bf r}\in B_1(\real^{N^2-1})$, the operator (\ref{formulaNxN}) does not necessarily describe a state. The reason for this is that $D({\bf r})$, to be a state, needs not only to be of unit trace and self-adjoint, but also positive semidefinite, which is not automatically guaranteed for an operator written as the real linear combination (\ref{formulaNxN}). Just to give an example, consider the unit vector ${\bf r}=(0,\dots,0,1)^T$. Then $D({\bf r})= {1\over N}\left(\mathbb{I} + c_N\Lambda_{N^2-1}\right)$, and according to (\ref{generatorsW}) $\Lambda_{N^2-1}=W_{N-1}$, so that $D({\bf r}) = -{N-2\over N} |b_N\rangle\langle b_N|+ {2\over N} \sum_{j=1}^{N-1} |b_j\rangle\langle b_j|$, which for $N\geq 3$ is clearly a matrix with a strictly negative eigenvalue $-{N-2\over N}$, and therefore cannot be positive semidefinite. 

What is important to observe is that although $B_1(\real^{N^2-1})$ is only partially filled with states, and that the shape of the region containing the states is rather complex, it is however a \emph{closed convex} region. This follows from the well-known fact that a convex linear combination of operator-states is again an operator-state. Thus, if ${\bf r} = a_1\, {\bf r}_1 + a_2\, {\bf r}_2$, with $a_1+a_2=1$, $a_1,a_2\geq0$, and ${\bf r}_1$ and ${\bf r}_2$ are representative of two operator-states, from (\ref{formulaNxN}) we immediately obtain that $D({\bf r})=a_1D({\bf r}_1)+a_2D({\bf r}_2)$ is again an operator-state, being a convex linear combination of operator states . We thus conclude that a vector ${\bf r}$, which is a convex linear combination of two vectors ${\bf r}_1$ and ${\bf r}_2$, representative of states, is also representative of a bona fide state $D({\bf r})$. 

We now explain how measurements can be represented within $B_1(\real^{N^2-1})$, so generalizing the previous two-dimensional measurement model. We start by observing that also in the general $N$-dimensional situation a vector-state $|\psi\rangle$ can always be represented, in a measurement context, by a one-dimensional projection operator $P_\psi = |\psi\rangle\langle\psi|$, as the only information one looses in the passage from $|\psi\rangle$ to $P_\psi$ is the global phase of the former. To show this, we write: $|\psi\rangle= \sum_{i=1}^N b_i |b_i\rangle$ and $P_\psi =\sum_{i,j=1}^N b_i b_j^* |b_i\rangle\langle b_j|$, so that $\langle b_k|P_\psi|b_\ell\rangle=b_kb_\ell^*$. For $k=\ell=1$, we have $\langle b_1|P_\psi|b_1\rangle=|b_1|^2$. Assuming that $b_1\neq 0$ (if this is not the case we can reason with $b_2$, and so on), we can always fix the global phase of $|\psi\rangle$ so as to have $b_1>0$. Then $b_1=\langle b_1|P_\psi|b_1\rangle^{1\over 2}$, and we obtain: $b_\ell=\langle b_\ell|P_\psi|b_1\rangle \langle b_1|P_\psi|b_1\rangle^{-{1\over 2}}$, $\ell =1,\dots,N$. Thus, apart a global phase factor, we can fully reconstruct $|\psi\rangle$ from $P_\psi$.

Our goal is to explain how the values of the transition probabilities ${\cal P}(P_\psi\to P_{a_i}) = {\rm Tr}\, P_\psi P_{a_i} =|\langle a_i|\psi\rangle|^2$ can be derived by means of a hidden-measurement mechanism, where $P_{a_i}= |a_i\rangle\langle a_i|$, and the orthonormal vectors $|a_1\rangle,\dots,|a_N\rangle$ are the eigenvectors of the spectral decomposition of an arbitrary observable $A=\sum_{i=1}^N a_i P_{a_i}$, which for the moment we shall assume to be non-degenerate (the eigenvalues $a_i$ are all distinct). 

Let ${\bf n}$ be the unit vector representative of $P_\psi\equiv P({\bf n})$, and ${\bf n}_i$ the unit vectors representative of $P_{a_i}\equiv P({\bf n}_i)$, $i=1,\dots,N$. We then have: 
\begin{eqnarray}
\lefteqn{{\cal P}(P_\psi \to P_{a_i}) = {\rm Tr}\, P({\bf n}) P({\bf n}_i) = {\rm Tr}\, {1\over N^2}\left(\mathbb{I} +c_N\, {\bf n}\cdot\mbox{\boldmath$\Lambda$}\right)\left(\mathbb{I} + c_N\,{\bf n}_i\cdot\mbox{\boldmath$\Lambda$}\right)}\nonumber\\
&=& {\rm Tr}\, {1\over N^2}\left[\mathbb{I} + c_N\,({\bf n}\cdot\mbox{\boldmath$\Lambda$}+{\bf n}_i\cdot\mbox{\boldmath$\Lambda$}) +c_N^2\,({\bf n}\cdot\mbox{\boldmath$\Lambda$})({\bf n}_i\cdot\mbox{\boldmath$\Lambda$})\right] = {1\over N} +{c_N^2\over N^2}\,{\rm Tr}\,({\bf n} \cdot\mbox{\boldmath$\Lambda$})\, ({\bf n}_i\cdot\mbox{\boldmath$\Lambda$})\nonumber\\
&=& {1\over N} \left[1+ (N-1)\,{\bf n}\cdot {\bf n}_i\right]={1\over N} \left[1+ (N-1)\,\cos\theta\right],
\label{transitiongeneralNxN}
\end{eqnarray}
where $\theta\equiv \theta({\bf n},{\bf n}_i)$ is the angle between ${\bf n}$ and ${\bf n}_i$. The above formula is clearly a generalization of (\ref{transition-probabilities2}), and we can use it to gain some knowledge on how the observables are represented within the unit ball $B_1(\real^{N^2-1})$. 

If the initial state is an eigenstate $P_{a_j}$, then we know that ${\cal P}(P_{a_j} \to P_{a_i}) = \delta_{ji}$. It follows from (\ref{transitiongeneralNxN}) that $\cos\theta({\bf n}_j,{\bf n}_i)=-{1\over N-1}$, that is: $\theta({\bf n}_j,{\bf n}_i) = \theta_N\equiv \cos^{-1} (-{1\over N-1})$, for all $i\neq j$. This means that the $N$ unit vectors ${\bf n}_i$, representative of the eigenstates $P_{a_i}$, $i=1\dots,N$, are the vertices of a $(N-1)$-dimensional \emph{simplex} $\triangle_{N-1}$, inscribed in the unit ball, with edges of length $\|{\bf n}_i -{\bf n}_j\| = \sqrt{2(1-\cos\theta_N)} = \sqrt{2N\over N-1}$. Also, considering that $\triangle_{N-1}$ is a convex set of vectors, it immediately follows that all points contained in it are representative of operator-states, in accordance with the fact that the states in $B_1(\mathbb{R}^{N^2-1})$ form a closed convex subset. 

For $N=2$, $\theta_1 = \pi$, and $\triangle_{1}$ is a line segment of length $2$ inscribed in a $3$-dimensional ball, as we have seen in the previous section. For $N=3$, $\theta_2 = {\pi\over 3}$, and $\triangle_{2}$ is an equilateral triangle of area ${3\sqrt{3}\over 4}$. For $N=4$, $\theta_3 \approx 0.6\,\pi$, and $\triangle_{3}$ is a tetrahedron of volume ${1\over 3}({4\over 3})^{3\over 2}$. For $N=5$, $\theta_4 \approx 0.58\,\pi$, and $\triangle_{4}$ is a pentachoron; and so on. 

We can write (\ref{transitiongeneralNxN}) in a more compact and simple form by expressing ${\bf n}$ as the sum: ${\bf n} = {\bf n}^\perp + {\bf n}^\parallel$, where ${\bf n}^\parallel$ is the vector obtained by orthogonally projecting ${\bf n}$ onto $\triangle_{N-1}$. Since by definition ${\bf n}^\perp\cdot {\bf n}_i =0$, $i=1,\dots, N$, (\ref{transitiongeneralNxN}) becomes ${\cal P}(P_\psi \to P_{a_i}) = {1\over N} \left[1+ (N-1)\,{\bf r}^\parallel\cdot {\bf n}_i\right]$. Also, as ${\bf n}^\parallel\in\triangle_{N-1}$, by definition of a simplex it can be uniquely written as a convex linear combination of the $N$ vertex vectors ${\bf n}_i$:
\begin{equation}
{\bf n}^\parallel =\sum_{i=1}^{N} n^\parallel_i \,{\bf n}_i, \quad \sum_{i=1}^{N} n^\parallel_i=1, \quad n^\parallel_i\geq 0, \,\, i=1\dots,N.
\label{n-parallelexpansion}
\end{equation}
Since ${\bf n}_i \cdot {\bf n}_j= -{1\over N-1}$, for $i\neq j$, we have: ${\bf n}^\parallel \cdot {\bf n}_i ={1\over N-1}(Nr^\parallel_i -1)$, and therefore (\ref{transitiongeneralNxN}) simply becomes:
\begin{equation}
{\cal P}(P_\psi \to P_{a_i}) = n^\parallel_i.
\label{trans-general}
\end{equation}

In the $N=2$ case we have seen that the measurement results from the interaction between an abstract point particle representative of the state of the entity and an attractive, elastic and unstable substance uniformly filling the one-dimensional simplex $\triangle_{1}$, representative of the measurement context. In the same way, the measurement context associated with a general $N$-dimensional observable consists of a $(N-1)$-dimensional simplex $\triangle_{N-1}$, uniformly filled with an attractive, elastic and unstable substance, with the point particle representative of the state that first orthogonally ``falls'' onto it and then is drawn to one of its apex points in an unpredictable way, as a result of the disintegration and collapse of said substance. 

\begin{figure}[!ht]
\centering
\includegraphics[scale =.7]{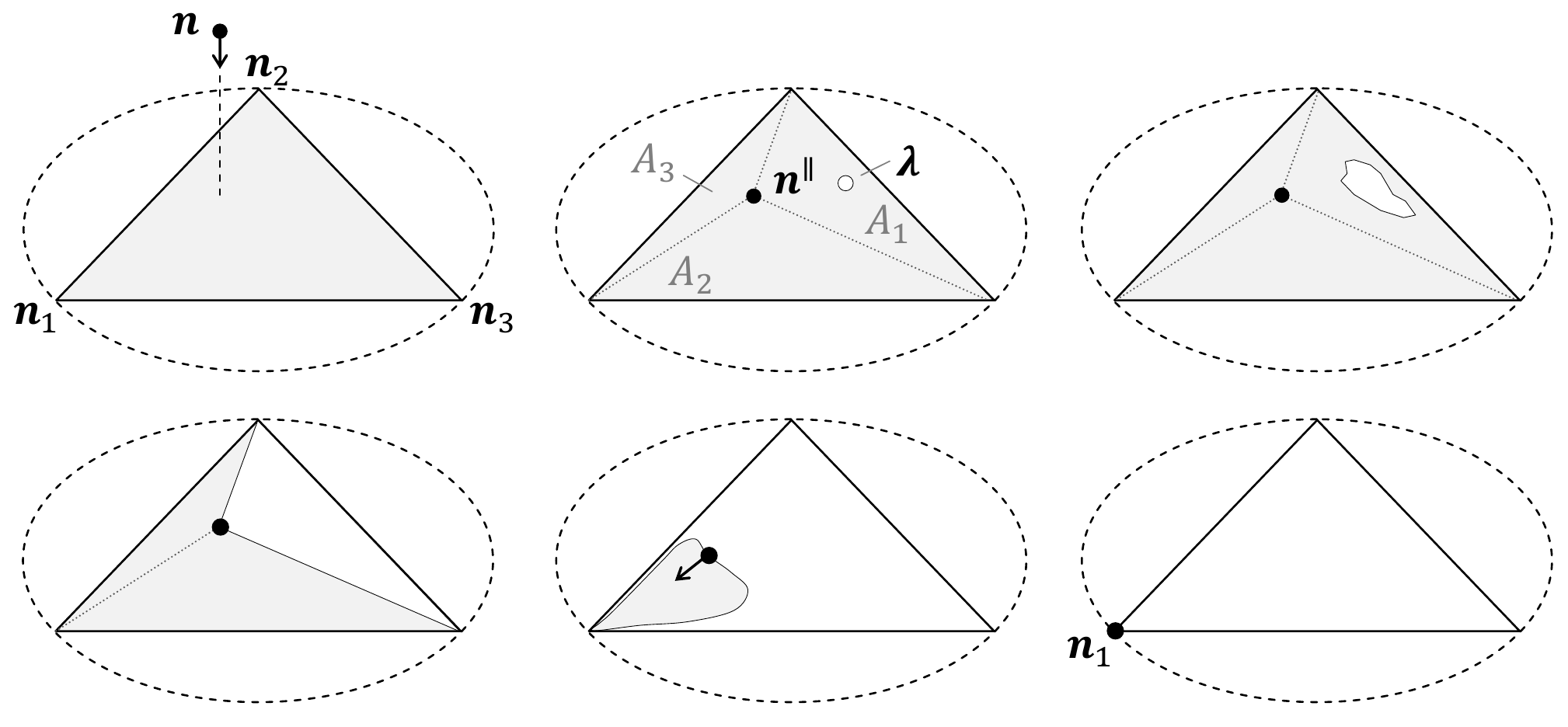}
\caption{The unfolding of a measurement process with three distinguishable outcomes: ${\bf n}_1$, ${\bf n}_2$ and ${\bf n}_3$. The point particle representative of the state, initially located in ${\bf{n}}$, deterministically approaches the triangular elastic membrane along an orthogonal path, reaching the on-membrane point ${\bf n}^\parallel$, defining in this way three subregions $A_1$, $A_2$ and $A_3$. The membrane then disintegrates at some unpredictable point \mbox{\boldmath$\lambda$}, producing the complete collapse of the associated subregion, here $A_1$, causing it to lose two of its anchor points and drawing in this way the point particle to its final location, here ${\bf{n}}_1$.
\label{spinmachine3d}}
\end{figure}

To see how all this works, we will only describe here, for simplicity, the $N=3$ situation, as the general situation proceeds according to the same logic and is therefore a straightforward generalization~\cite{AertsSassoli2014c}. So, the measurement context is in this case represented by a $2$-dimensional triangular elastic membrane inscribed in a 8-dimensional ball, and the three possible outcomes are the eigenstates associated with its three vertex vectors ${\bf n}_1$, ${\bf n}_2$ and ${\bf n}_3$. If the initial, pre-measurement state $P({\bf n})$ is associated with a unit vector ${\bf n}\in B_1(\real^{8})$, the entity proceeds first with a deterministic movement ${\bf r}(\tau) = (1-\tau)\,{\bf n} + \tau\, {\bf n}^\parallel$, $\tau \in [0,1]$, which brings the state of the entity from the point at the surface ${\bf r}(0) = {\bf n}$, to the on-membrane point ${\bf r}(1)={\bf n}^\parallel$, along a path that is orthogonal to $\triangle_{2}$ (see Fig.~\ref{spinmachine3d}).

Once the particle has reached its on-membrane position ${\bf n}^\parallel$, it defines three different triangular subregions $A_1$, $A_2$ and $A_3$, delineated by the line segments connecting the particle's position with the three vertex points (see Fig.~\ref{spinmachine3d}). One should think of these line segments as ``tension lines'' altering the functioning of the membrane, in the sense of making it less easy to disintegrate along them. Then, at some moment, the membrane disintegrates, at some unpredictable point \mbox{\boldmath$\lambda$}, belonging to one of these three subregions. The disintegration then propagates inside that specific subregion, but not into the other two subregions, because of the presence of the tension lines. This causes the two anchor points of the disintegrating subregion to tear away, producing the detachment of the membrane, which being elastic contracts towards the only remaining anchor point, drawing to that position also the particle that is attached to it, which in this way reaches its final destination (state), corresponding to the outcome of the measurement (see Fig.~\ref{spinmachine3d}). 

Reasoning in the same way as we did in the $N=2$ case, it is clear that the transition probability ${\cal P}(P_\psi\to P_{a_i})$ is given by the probability ${\cal P}(\mbox{\boldmath$\lambda$}\in A_i)$ that the disintegration point \mbox{\boldmath$\lambda$} belongs to region $A_i$. Considering that $\triangle_{2}$ is an equilateral triangle of area ${3\sqrt{3}\over 4}$, we thus have ${\cal P}(\mbox{\boldmath$\lambda$}\in A_i)= {4\over 3\sqrt{3}}\mu(A_i)$. Let us consider for instance $A_1$. It is a triangle with vertices ${\bf n}_2$, ${\bf n}_3$ and ${\bf n}^\parallel = \sum_{i=1}^3 n^\parallel_i {\bf n}_i$. Using the explicit coordinates of the three vertices of $A_1$, we can easily calculate its area. For this, we can use a system of coordinate directly in the plane of the triangle, such that ${\bf n}_2=(0,1)\equiv (x_1,x_2)$, ${\bf n}_3=({\sqrt{3}\over 2},-{1\over 2})\equiv (y_1,y_2)$, and ${\bf n}_1=(-{\sqrt{3}\over 2},-{1\over 2})\equiv (z_1,z_2)$. To calculate the area we can then use the general formula: $\mu(A_1)={1\over 2}\left|-y_1x_2 +z_1x_2+x_1y_2-z_1y_2-x_1z_2+y_1z_2 \right|$. After a calculation without difficulties one finds, using (\ref{n-parallelexpansion}): $\mu(A_1)={3\sqrt{3}\over 4}n^\parallel_1$. Doing similar calculation for $A_2$ and $A_3$, one obtains that $\mu(A_i)={3\sqrt{3}\over 4}n^\parallel_i$, $i=1,2,3$, so that ${\cal P}(\mbox{\boldmath$\lambda$}\in A_i)= {4\over 3\sqrt{3}}\mu(A_i)=n^\parallel_i$, $i=1,2,3$, in accordance with the predictions of the quantum mechanical Born rule (\ref{trans-general}).

The hidden-measurement membrane mechanism that we have described for the $N=2$ case in the previous section, and for the $N=3$ case in the present section, generalizes in a natural way to an arbitrary number of dimensions $N$, and we refer the reader to~\cite{AertsSassoli2014c} for all the mathematical details. The membrane mechanism can also be used to describe measurements of degenerate observables. Then, the subregions associated with the degenerate eigenvalues are fused together and form bigger composite subregions, so that when the initial disintegration point \mbox{\boldmath$\lambda$} takes place inside one of them, the process draws the particle not to a vertex point of $\triangle_{N-1}$, but to one of its sub-simplexes. The collapse of the elastic substance remains however compatible with the predictions of the \emph{L\"uders-von Neumann projection formula}, but to complete the process the particle also has to re-emerge from the sub-simplex region, to deterministically reach its final position, at the surface of the unit ball. 

In other words, in the general situation a measurement is to be understood as a tripartite process formed by (1) an initial deterministic \emph{decoherence-like} process, corresponding to the particle reaching the ``on-membrane region of potentiality;'' (2) a subsequent indeterministic \emph{collapse-like} process, corresponding to the disintegration of the elastic substance filling the simplex, with the particle being drawn to some of its peripheral points; and (3) a possible final deterministic \emph{purification-like} process, bringing again the particle to a unit distance from the center of the sphere~\cite{AertsSassoli2014c}.

\section{Spin eigenstates and space directions}
\label{Spin and space}

In the previous two sections we have shown that not only the standard Bloch sphere representation of two-state systems, like spin-${1\over 2}$ entities, can be generalized to describe general $N$-state systems, like spin-$s$ entities with $s={N-1\over 2}$, but also that it can be extended -- by means of a hidden-measurement mechanism -- to include a full representation of all possible measurements. We have therefore answered the question regarding the nature of the interaction between the quantum and classical theaters (an answer which constitutes a solution to the measurement problem). In this section, we want to answer the second question we have addressed in the Introduction, regarding the general relation between spin eigenstates and space directions, by exploiting the descriptive power of the extended Bloch model. 

We have already observed that the classical-like image of a quantum spin, as a set of undetermined classical angular momentum vectors lying on a cone, is inconsistent, but we have also seen that for a spin-${1\over 2}$ entity there is a direct one-to-one correspondence between spin eigenstates and directions of space. These directions, however, are not anymore those characterizing classical angular momenta, but those for which the states produce a predetermined outcome, in spin measurements along those same directions. 

Now, even beyond the $s={1\over 2}$ case, eigenstates will continue to be associated with space directions, as is clear that they are the eigenstates of spin observables that can only be defined by specifying a space direction (the orientation of the Stern-Gerlach apparatus). However, we cannot expect anymore  each eigenstate to be entirely characterized by a single space direction, as is clear that more than a single eigenvalue can be observed for each direction. In addition to that, we know that in the extended Bloch model, states are not anymore represented by three-dimensional vectors, beyond the special spin-${1\over 2}$ situation. So, how do spin eigenstates relate, in general, to the directions of our 3-dimensional space, that is, to the directions defining the spin observables for which they are the eigenstates? The answer to this question is contained in the following:
\\

\noindent {\bf Proposition 1}. \emph{Given a spin-$s$ entity, with $s={N-1\over 2}$, $N\geq 2$, and a spin observable $S_{\hat{\bf n}}$, oriented along the space direction $\hat{\bf n}$, then to its $N=2s+1$ eigenstates $|\hat{\bf n},\mu\rangle\in \compl^{N}$, $\mu =-s,\dots,s$, represented in the extended Bloch sphere by the unit vectors ${\bf n}_{\mu}\in B_1(\real^{N^2-1})$, $\mu =-s,\dots,s$, we can associate a unit vector ${\bf v}\in B_1(\real^{N^2-1})$, defined by:
\begin{equation}
{\bf v} =d_N \sum_{\mu =-s}^{s}\mu \, {\bf n}_{\mu},\quad d_N={1\over N}\sqrt{12\over N+1},
\label{vectorv}
\end{equation}
such that when the three-dimensional unit vector $\hat{\bf n}$ runs through all the possible directions of the Euclidean space, ${\bf v}$ equally spans, in an isomorphic way, all the points at the surface of a $3$-dimensional sub-ball of $B_1(\real^{N^2-1})$.
}\\

\begin{figure}[!ht]
\centering
\includegraphics[scale =.57]{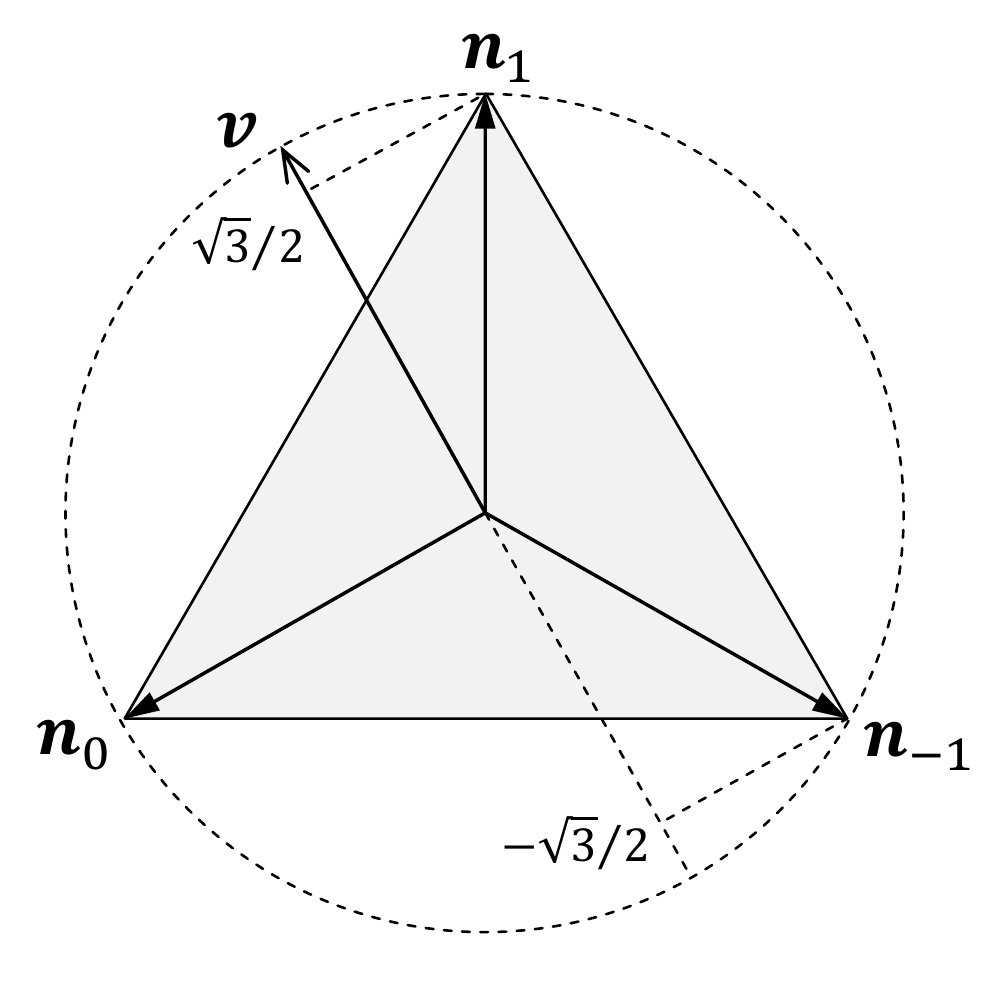}
\caption{The 2-simplex of a spin-1 measurement, inscribed in a 8-dimensional ball, with the three eigenvectors ${\bf n}_{-1},{\bf n}_{0}$ and ${\bf n}_{1}$, and the associated ``space'' vector ${\bf v} ={1\over\sqrt{3}}({\bf n}_{1}-{\bf n}_{-1})$. 
\label{vectorvfigure}}
\end{figure}

To prove the proposition, we write $S_{\hat{\bf n}}={\bf S}\cdot \hat{\bf n}=S_1n_1 + S_2n_2 + S_3n_3$, where $\hat{\bf n}=n_1 \hat{\bf x}_1+n_2 \hat{\bf x}_2 +n_3 \hat{\bf x}_3$, $S_i={\bf S}\cdot \hat{\bf x}_i$, and the $\hat{\bf x}_i$, $i=1,2,3$, are the orthonormal vectors defining the canonical basis of the 3-dimensional Euclidean space (we have introduced the ``hat notation'' to distinguish 3-dimensional vectors from $(N^2-1)$-dimensional ones). We define $P({\bf n}_\mu)=|\hat{\bf n},\mu\rangle\langle \hat{\bf n},\mu|={1\over N}(\mathbb{I}+c_N \,{\bf n}_\mu \cdot \mbox{\boldmath$\Lambda$})$, and we also consider the eigenstates $P({\bf n}'_\nu)=|\hat{\bf n}',\nu\rangle\langle \hat{\bf n}',\nu|={1\over N}(\mathbb{I}+c_N \,{\bf n}'_\mu \cdot \mbox{\boldmath$\Lambda$})$, of $S_{\hat{\bf n}'}={\bf S}\cdot\hat{\bf n}'$, and the associated vector ${\bf v}'=d_N \sum_{\nu =-s}^{s}\nu \, {\bf n}'_{\nu}$. We want to calculate the scalar product: 
\begin{equation}
{\bf v}\cdot {\bf v}'=d_N^2\sum_{\mu,\nu =-s}^{s}\mu\nu \, {\bf n}_{\mu}\cdot {\bf n}'_{\nu}.
\label{vectorv2}
\end{equation}
Using the spectral decompositions $S_{\hat{\bf n}}=\sum_{\mu =-s}^{s}\hbar\mu \, P({\bf n}_\mu)$ and $S_{\hat{\bf n}'}=\sum_{\nu =-s}^{s}\hbar\nu \, P({\bf n}'_\nu)$, we have: 
\begin{eqnarray}
{\rm Tr}\, S_{\hat{\bf n}}S_{\hat{\bf n}'} &=& \hbar^2\sum_{\mu,\nu =-s}^{s}\mu\nu \, {\rm Tr}\,P({\bf n}_\mu)P({\bf n}'_\nu)={\hbar^2\over N^2} \sum_{\mu,\nu =-s}^{s}\mu\nu \, {\rm Tr}\, (\mathbb{I}+c_N \,{\bf n}_\mu \cdot \mbox{\boldmath$\Lambda$})(\mathbb{I}+c_N \,{\bf n}'_\nu \cdot \mbox{\boldmath$\Lambda$})\nonumber\\
&=& {\hbar^2 c_N^2\over N^2} \sum_{\mu,\nu =-s}^{s}\mu\nu \, {\rm Tr}\, ({\bf n}_\mu \cdot \mbox{\boldmath$\Lambda$}) ({\bf n}'_\nu \cdot \mbox{\boldmath$\Lambda$})= {2\hbar^2 c_N^2\over N^2} \sum_{\mu,\nu =-s}^{s}\mu\nu \, {\bf n}_\mu \cdot {\bf n}'_\nu,
\end{eqnarray}
where for the penultimate equality we have used the fact that the $\Lambda_i$ are traceless, and for the last equality that they obey ${\rm Tr}\,\Lambda_i\Lambda_j=2\delta_{ij}$. So, (\ref{vectorv2}) becomes: 
\begin{equation}
{\bf v}\cdot {\bf v}'= {N^2d_N^2\over 2\hbar^2 c_N^2} {\rm Tr}\, S_{\hat{\bf n}}S_{\hat{\bf n}'} ={1\over\hbar^2}{12\over (N-1)N(N+1)}{\rm Tr}\, S_{\hat{\bf n}}S_{\hat{\bf n}'}.
\label{vectorv3}
\end{equation}

To evaluate the right hand side of (\ref{vectorv3}), we observe that the three spin operators $S_1$, $S_2$ and $S_3$, associated with orthogonal directions of space, are orthogonal operators: ${\rm Tr}\, S_iS_j = 0$, if $i\neq j$. To see this, let $U$ be the unitary operator representing a rotation around axis-$j$, such that $U^\dagger S_i U =-S_i$. As we have $[U, S_j]=0$, from the cyclicity of the trace it follows that: ${\rm Tr}\, S_iS_j = {\rm Tr}\, U^\dagger S_iS_j U = {\rm Tr}\, U^\dagger S_iUS_j =-{\rm Tr}\, S_iS_j$, and therefore we find that ${\rm Tr}\, S_iS_j = 0$. Therefore: 
\begin{eqnarray}
{\rm Tr}\, S_{\hat{\bf n}}S_{\hat{\bf n}'} &=& {\rm Tr}\,\left({\bf S}\cdot \hat{\bf n}\right)\left({\bf S}\cdot \hat{\bf n}'\right)=\sum_{i,j=1}^3 n_in'_j {\rm Tr}\,S_iS_j= \sum_{i=1}^3 {\rm Tr}\,S_i^2 \, n_in'_i \nonumber\\
&=& {\hbar^2 s(s+1)\over 3}N \sum_{i=1}^3 n_in'_i = \hbar^2{(N-1)N(N+1)\over 12}\, \hat{\bf n}\cdot\hat{\bf n}'.
\label{vectorv4}
\end{eqnarray}
Finally, comparing (\ref{vectorv4}) with (\ref{vectorv3}), we obtain the equality: 
\begin{equation}
{\bf v}\cdot {\bf v}' = \hat{\bf n}\cdot\hat{\bf n}',
\label{scalarproduct}
\end{equation}
showing that the scalar product between the two vectors ${\bf v}=d_N \sum_{\nu =-s}^{s}\nu \, {\bf n}_{\nu}$ and ${\bf v}'=d_N \sum_{\nu =-s}^{s}\nu \, {\bf n}'_{\nu}$, belonging to $B_1(\real^{N^2-1})$, is isomorphic to the scalar product between the corresponding vectors $\hat{\bf n}$ and $\hat{\bf n}'$, belonging to $B_1(\real^{3})$, describing the spatial orientation of the associated spin measurement apparatuses.\\

The following remarks are in order. The vectors ${\bf v}$ are in a one-to-one correspondence with the space directions $\hat{\bf n}$. Therefore, we can say that they represent these directions within the quantum Blochean theater. We observe that: 
\begin{eqnarray}
{\bf v}\cdot {\bf n}_\nu &=& d_N\sum_{\mu=-s}^s \mu\, {\bf n}_\mu\cdot {\bf n}_\nu =d_N \left(\nu -{1\over N-1}\sum_{\mu\neq\nu}^s \mu\right) = d_N \left[\nu -{1\over N-1}\left(-\nu + \sum_{\mu=-s}^s\mu\right)\right]\\
&=&d_N \left(\nu +{\nu\over N-1}\right) = {Nd_N\over N-1}\,\nu ={1\over N-1}\sqrt{12\over N+1}\,\nu,
\end{eqnarray}
where for the second equality we have used ${\bf n}_\mu\cdot{\bf n}_\nu=-{1\over N-1}$, for $\mu\neq\nu$. For $s={1\over 2}$ ($N=2$), we have ${\bf v}\cdot {\bf n}_{\pm{1\over 2}}=\pm1$, i.e., ${\bf v}={\bf n}_{1\over 2}$, meaning that the spin eigenstates are entirely characterized by the space directions. On the other hand, for $s>{1\over 2}$ ($N>2$), the eigenvectors are not anymore oriented along the space directions ${\bf v}$ (we abusively say ``space directions,'' while it is clear that we should say, more precisely, ``representatives of the space directions $\hat{\bf n}$ within $B_1(\real^{N^2-1})$''). However, they all have a well-defined and fixed orientation with respect to ${\bf v}$, which depends on their eigenvalue and which is such that, when orthogonally projected onto ${\bf v}$, they give rise to equally spaced points (see Fig.~\ref{vectorvfigure} for the $s=1$ case), which are somehow reminiscent of the spots observed on the screen of a Stern-Gerlach apparatus, along the magnetic field direction. 

This doesn't mean, however, that when ${\bf v}$ spans a three-dimensional sub-ball of $B_1(\real^{N^2-1})$, correspondingly to as $\hat{\bf n}$ spans $B_1(\real^3)$, the vertex vectors ${\bf n}_\mu$ will also span the surfaces of $3$-dimensional sub-balls. Indeed, there are no unit sub-balls filled with states within $B_1(\real^{N^2-1})$ (see below).

The ``space'' vectors ${\bf v}$ have a simple expression in the $N=2$ and $N=3$ cases. Indeed, for $N=2$, we have: ${\bf v} = {1\over 2}({\bf n}_{1\over 2}-{\bf n}_{-{1\over 2}})$, and for $N=3$, we have: ${\bf v} = {1\over \sqrt{3}}({\bf n}_{1}-{\bf n}_{-1})$ (see Fig.~\ref{vectorvfigure}). In other terms, for elementary spin-${1\over2}$ fermions and spin-1 bosons, the space directions ${\bf v}$ are simply given by the difference of the two eigenvectors of extremal eigenvalues. For spin-${1\over 2}$ entities this is so because there are only two eigenvectors available, whereas for spin-1 entities it is because one of the eigenvectors is associated with a zero eigenvalue, and therefore cannot contribute to the sum (\ref{vectorv}). Note that the fact that ${\bf v}$ is given by the normalized difference of two eigenvectors, means that the associated edge of the measurement simplex is oriented like ${\bf v}$. 

Equation (\ref{scalarproduct}) immediately implies that ${\bf v}$ is a unit vector. One may wonder if it can be associated with a physical state. A priori, there is no reason for this, as a (normalized) superposition of states within the Bloch sphere will generally not be a state, beyond the $N=2$ case. And in fact, if $N>2$, ${\bf v}$ is never representative of a vector-state. To see this, let $P({\bf v})={1\over N}(\mathbb{I}+c_N \,{\bf v} \cdot \mbox{\boldmath$\Lambda$})$. If $P({\bf v})$ would be representative of a state, we would have ${\rm Tr}\,P({\bf v})P({\bf n}_\mu)\geq 0$, as is clear that the transition probability from a state to another state must always be a positive number. However: 
\begin{eqnarray}
{\rm Tr}\,P({\bf v})P({\bf n}_\mu) &=&{1\over N}[1+ (N-1) {\bf v}\cdot {\bf n}_\mu ]= {1\over N}\left[1+ (N-1) d_N\sum_{\nu=-s}^s {\bf n}_\nu\cdot{\bf n}_\mu \right] \nonumber\\
&=& {1\over N}\left[1+ (N-1) d_N(\mu +{1\over N-1}\,\mu )\right]={1\over N}\left[1+ \sqrt{12\over N+1}\,\mu\right],
\label{tracedips}
\end{eqnarray}
So, for the extremal value $\mu=-s=-{N-1\over 2}$, we have: 
\begin{equation}
{\rm Tr}\, P({\bf v})P({\bf n}_{-s}) = {1\over N}\left(1 -\sqrt{3(N-1)^2\over N+1}\right)= {1\over N}\left(1 -\sqrt{1+{(N-2)(3N-1)\over N+1}}\right),
\label{tracedips2}
\end{equation}
which means that ${\rm Tr}\, P({\bf v})P({\bf n}_{-s})<0$, if $N>2$, implying that ${\bf v}$ cannot be representative of a bona fide vector-state. In fact, and more generally, one can prove that in the $(N-1)$-dimensional subspace generated by eigenvectors ${\bf n}_\mu$, $\mu=-s,\dots,s$, these are the only \emph{unit} vectors that can be associated with physical states (see~\cite{AertsSassoli2014c} for a proof of this statement).

\section{Composite spin entities}
\label{Composition of spins}

In the present section we explore the relation between spin eigenstates and space directions in the case of a spin entity formed by two sub-entities. We have the following:\\

\noindent {\bf Proposition 2}. \emph{Given two spin entities, of spin $s_1={N_1-1\over 2}$ and $s_2={N_2-1\over 2}$, and the spin observables $S_{\hat{\bf n}}^{(1)}$ and $S_{\hat{\bf n}}^{(2)}$, oriented along the space direction $\hat{\bf n}$, acting on $\compl^{N_1}$ and $\compl^{N_2}$, respectively, and given the total spin observables $S_{\hat{\bf n}}= S_{\hat{\bf n}}^{(1)}\otimes \mathbb{I}^{(2)} + \mathbb{I}^{(1)}\otimes S_{\hat{\bf n}}^{(2)}$ and $S^2$, 
acting on $\compl^{N_1}\otimes \compl^{N_2}\equiv \compl^{N_1N_2}$, then, to the $N=N_1N_2 = (2s_1+1)(2s_2+1)$ eigenstates $|\hat{\bf n},s,\mu_s\rangle\in \compl^{N}$, such that $S_{\hat{\bf n}}|\hat{\bf n},s,\mu_s\rangle =\hbar\mu _s |\hat{\bf n},s,\mu_s\rangle$ and $S^2|\hat{\bf n},s,\mu_s\rangle =\hbar^2 s(s+1) |\hat{\bf n},s,\mu_s\rangle$,
$s = |s_1-s_2|,\dots, s_1+s_2$, $\mu_s =-s,\dots,s$, represented in the extended Bloch sphere by the unit vectors ${\bf n}_{s,\mu_s}\in B_1(\real^{N^2-1})$, we can associate a unit vector ${\bf v}\in B_1(\real^{N^2-1})$, defined by:
\begin{equation}
{\bf v} =d_{N_1,N_2} \sum_{s = |s_1-s_2|}^{s_1+s_2}\sum_{\mu _s=-s}^{s}\mu _s\, {\bf n}_{s,\mu_s},\quad d_{N_1,N_2}=\sqrt{12(N-1)\over NN_1N_2 (N_1^2 + N_2^2-2)},
\label{vectorv2-bis}
\end{equation}
such that when the three-dimensional unit vector $\hat{\bf n}$ runs through all the possible directions of the Euclidean space, ${\bf v}$ equally spans, in an isomorphic way, all the points at the surface of a $3$-dimensional sub-ball of $B_1(\real^{N^2-1})$.
}\\

The proof follows almost identically that of \emph{Proposition 1}. We observe that: 
\begin{equation}
S_i^2 = (S_i^{(1)}\otimes \mathbb{I}^{(2)} + \mathbb{I}^{(1)}\otimes S_i^{(2)})(S_i^{(1)}\otimes \mathbb{I}^{(2)} + \mathbb{I}^{(1)}\otimes S_i^{(2)}) = ({S_i^{(1)}})^2\otimes \mathbb{I}^{(2)}+\mathbb{I}^{(1)}\otimes ({S_i^{(2)}})^2 + 2 S_i^{(1)}\otimes S_i^{(2)},
\end{equation}
and since the trace of the tensor product of two matrices is the product of their traces, we have ${\rm Tr}\, S_i^{(1)}\otimes S_i^{(2)}= {\rm Tr}\, S_i^{(1)}{\rm Tr}\, S_i^{(2)}=0$, so that:
\begin{eqnarray}
{\rm Tr}\, S_i^2 &=& {\rm Tr}\, ({S_i^{(1)}})^2 {\rm Tr}\,{I}^{(2)} + {\rm Tr}\,{I}^{(1)}{\rm Tr}\,({S_i^{(2)}})^2 = {\hbar^2 s_1(s_1+1)\over 3}N_1N_2 + N_1N_2{\hbar^2 s_2(s_2+1)\over 3}\nonumber\\
&=& {\hbar^2N_1N_2\over 12}\left(N_1^2+ N_2^2-2\right).
\end{eqnarray}
Instead of (\ref{vectorv3}), we thus have: 
\begin{equation}
{\bf v}\cdot {\bf v}'= {N^2d_{N_1,N_2}^2\over 2\hbar^2 c_N^2} {\rm Tr}\, S_{\hat{\bf n}}S_{\hat{\bf n}'} ={1\over\hbar^2}{12\over N_1N_2(N_1^2+N_2^2-2)}{\rm Tr}\, S_{\hat{\bf n}}S_{\hat{\bf n}'},
\label{vectorv3-bis}
\end{equation}
and instead of (\ref{vectorv4}), we have:
\begin{equation}
{\rm Tr}\, S_{\hat{\bf n}}S_{\hat{\bf n}'}= \sum_{i=1}^3 {\rm Tr}\,S_i^2 \, n_in'_i = {\hbar^2N_1N_2\over 12}\left(N_1^2+ N_2^2-2\right)\, \hat{\bf n}\cdot\hat{\bf n}'.
\label{vectorv4-bis}
\end{equation}
Comparing (\ref{vectorv4-bis}) with (\ref{vectorv3-bis}), we then obtain  the isomomorphism (\ref{scalarproduct}), that is: ${\bf v}\cdot {\bf v}'=\hat{\bf n}\cdot\hat{\bf n}'$.\\

The following remarks are in order. In the previous section we have observed that, for non-composite spin entities, the $s={1\over 2}$ and $s=1$ cases were special, in the sense that the difference of the two extremal eigenstates (corresponding to one of the edges of the measurement simplex) is oriented like ${\bf v}$. When we combine two spin-${1\over 2}$ entities ($s_1=s_2={1\over 2}$), we have four eigenstates, but two of them have zero eigenvalue: the singlet state ${\bf n}_{0,0}$ and the triplet state ${\bf n}_{1,0}$, which therefore cannot contribute to the sum (\ref{vectorv2-bis}). This means that also in this case ${\bf v}$ is simply given by the difference of the two extremal spin states: ${\bf v}= \sqrt{3\over 8}({\bf n}_{1,1}-{\bf n}_{1,-1})$.

We also observe that: ${\bf v}\cdot {\bf n}_{s,\mu_s} = {Nd_{N_1N_2}\over N-1}\,\mu_s$, which means that the eigenvector-states ${\bf n}_{s,\mu_s}$, when orthogonally projected onto ${\bf v}$, give rise also in this case to equally spaced points. However, since there are now degenerate eigenvalues, although all the vectors ${\bf n}_{s,\mu_s}$ are oriented differently within $B_1(\real^{N^2-1})$, as they are the vertex vectors of a $(N-1)$-simplex, those associated with the same eigenvalue will produce the same spot, when projected onto the ${\bf v}$-axis (see Fig.~\ref{correspondence2}).

\section{Connecting classical and quantum elements of reality}
\label{Connecting elements of reality}

In the introductory section, we have explained that because of our human condition within reality, we can only explore its content from the perspective of our Euclidean theater. Therefore, our situation is in a sense similar to that described by the philosopher \emph{Plato} in his famous allegory of the cave. However, Plato opposed the material world known to us through our senses (described as the shadows on the cave's wall) to a world of immutable ``ideas,'' possessing an ultimate reality. Different from Plato, the view we bring forward in this article is that although our Euclidean theater is like a Plato's cave, being the expression of a limited perspective, different caves possibly exist, corresponding to the different vantage points on reality that in principle can be adopted.

What we have called the ``quantum theater'' is in fact nothing but another cave, one not directly inhabited by us humans, and we could generally say that the whole of reality is a construction about the different possible caves (or theaters, we use these two terms here as synonyms). Our work, as investigators and participators of reality, is then not only that of identifying the content of the cave we inhabit, and of the other possible caves, but also the relations (the partial morphisms) that exist between their different elements of reality. This means that we don't necessarily ask reality to be fully contained in a single fundamental theater, although of course we cannot exclude that such an ultimate stage couldn't one day be identified. 

We neither cannot exclude that there may be elements of reality belonging to one cave that cannot find a direct correspondence in another cave. We already know that certain caves are, in a sense, bigger than others. For instance, most physicists believe that the quantum cave should contain the classical one, but this cannot be taken for granted, as we know that there are severe structural shortcomings in the quantum representation. For example, in the standard quantum formalism, entities that are fully separated in experimental terms cannot be described, whereas there are plenty of separated entities in the classical theater~\cite{Aerts1982b, Aerts1984, Aerts1999b, Aerts2000}. Also, there are elements of reality which appear to be of a genuine intermediate nature, which cannot be represented neither in the classical nor in the quantum caves, as they belong to a truly intermediate representation, which so far has not received much attention~\cite{Aerts1998b, Aerts1999, AertsSassoli2014a,AertsSassoli2014b}. And it may very well be that the many failed tentative to unify gravitational and quantum elements of reality, within a same consistent big theater, could be due to the fact that, for structural reasons, a single ``quantum \& gravitational'' theater cannot be constructed. 

Considering our description of the measurement process, we should also ponder that there is an additional difficulty to take into consideration, when different theaters, or caves, are put in relation with each other. Plato, in his allegory, used the concept of ``shadows,'' that the entities outside of our human cave casted onto its lower dimensional walls. The observation of shadows, however, implicitly assumes a classical modality of observation, only involving a discovery aspect. This is the modality of observation that physicists know to apply, at least in principle, when classical entities are observed by means of other classical entities. But as we have seen in Sec.~\ref{The extended Bloch sphere} and Sec.~\ref{General measurement}, when quantum entities are observed, a creation aspect is also involved, i.e., the observation can also partly produce the observed property. 

This last remark is crucial in order to understand how the correspondences between elements of reality belonging to different caves should be understood. In that respect, let us make precise that what we mean here by the notion of ``element of reality'' is exactly what was meant by \emph{Einstein}, \emph{Podolsky} and \emph{Rosen}, in their famous 1935 article~\cite{Einstein1935}. An element of reality is a \emph{state of prediction}: a property of an entity that we know is \emph{actual}, in the sense that, should we decide to observe it (i.e., to test its actuality), the outcome of the observation would be certainly successful. So, what we have to consider is that the way we observe a classical entity is different form the way we observe a quantum one. More specifically, and coming back now to the example of spins, which more particularly concerns us in this article, when we compare a quantum spin with a classical spin, i.e., with an angular momentum understood as a ``state of rotation in the three-dimensional Euclidean space,'' we need not to forget that their elements of reality are operationally defined in a different way. Indeed, for a quantum spin an observation can be invasive, unless the entity is precisely in an eigenstate of the observation in question, whereas for a classical spin the observation is always non-invasive, independently of the state of the entity. 

In other terms, when we look for those elements of reality belonging to the quantum cave that can be put in a correspondence with the elements of reality belonging to the classical one, only eigenstates can be considered, and more precisely eigenstates in association with their specific measurements. This because only for eigenstates the observational process is of the pure discovery kind, as is the case for all classical elements of reality, and of course it would make no sense to try to compare quantum \emph{potential} properties, described by \emph{superpositions} of eigenstates, with classical \emph{actual} properties, as superpositions of eigenstates truly describe elements of the quantum cave that have no direct counterpart with what is contained in the classical one.

\begin{figure}[!ht]
\centering
\includegraphics[scale =0.57]{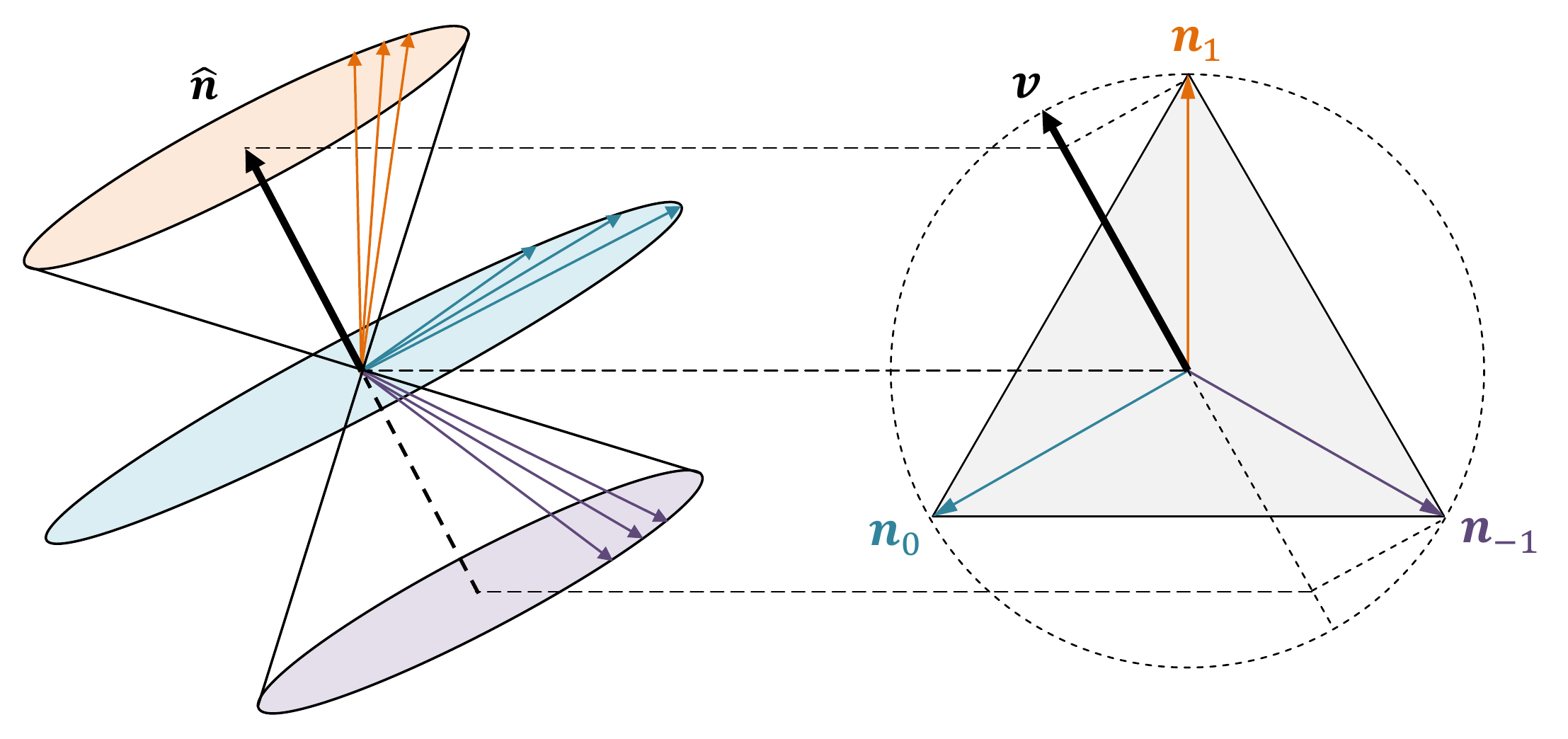}
\caption{The correspondence between the 3-dimensional classical theater and the 8-dimensional quantum theater, for a (non-composite) spin-1 entity. To each eigenstate in the latter, associated with vector ${\bf v}$, there is a cone of angular momentum vectors in the former, associated with vector ${\hat{\bf n}}$. These cones, however, can only represent the spin eigenstates for the measurements of the two commuting observables $S^2$ and $S_{\hat{\bf n}}$. For other observables, there are no more correspondences between the effects of observations on the quantum and classical spin entities. Note also that, contrary to what the figure may suggest, the plane containing the measurement triangle is not part of the 3-dimensional sub-sphere in which ${\bf v}$ moves, isomorphically to ${\hat{\bf n}}$. 
\label{correspondence}}
\end{figure}

This is how we have to understand the correspondence between classical spins (angular momenta) and quantum spin eigenstates, as expressed in our \emph{Proposition 1}. Consider Fig.~\ref{spinfamily}, and more particularly the second structure from the left, representing the three eigenstates of a spin-1 entity. Each one of the three cones represents a collection of classical angular momenta of length $\hbar \sqrt{2}$ (compatibly with $S^2=\hbar^2 2\, \mathbb{I}$), whose spatial orientations are consistent with the elements of reality of the corresponding quantum spin eigenstates (since when projected onto the central axis they give rise to vectors of length $\hbar$, 0 and $-\hbar$, respectively), forming an equilateral triangle in the 8-dimensional Bloch sphere. The central axis of the cone corresponds to the direction of observation $\hat{\bf n}$, which in the case of a quantum measurement is the orientation of the Stern-Gerlach magnetic field. As we have shown, this direction of space can also be identified within the extended Bloch sphere, and corresponds to the (here 8-dimensional) vector ${\bf v}$ (see Fig.~\ref{correspondence}). This means that when we move the cones in the Euclidean space, the simplex of eigenstates will move accordingly, in the generalized Blochean space of directions, in order to preserve the correspondence between the different elements of reality. The correspondence, however, remains meaningful only if a single measurement is considered at a time, as is clear that eigenstates give rise to predictable outcomes only with respect to a given measurement (and to those other measurements that are compatible with it). 
\begin{figure}[!ht]
\centering
\includegraphics[scale =0.57]{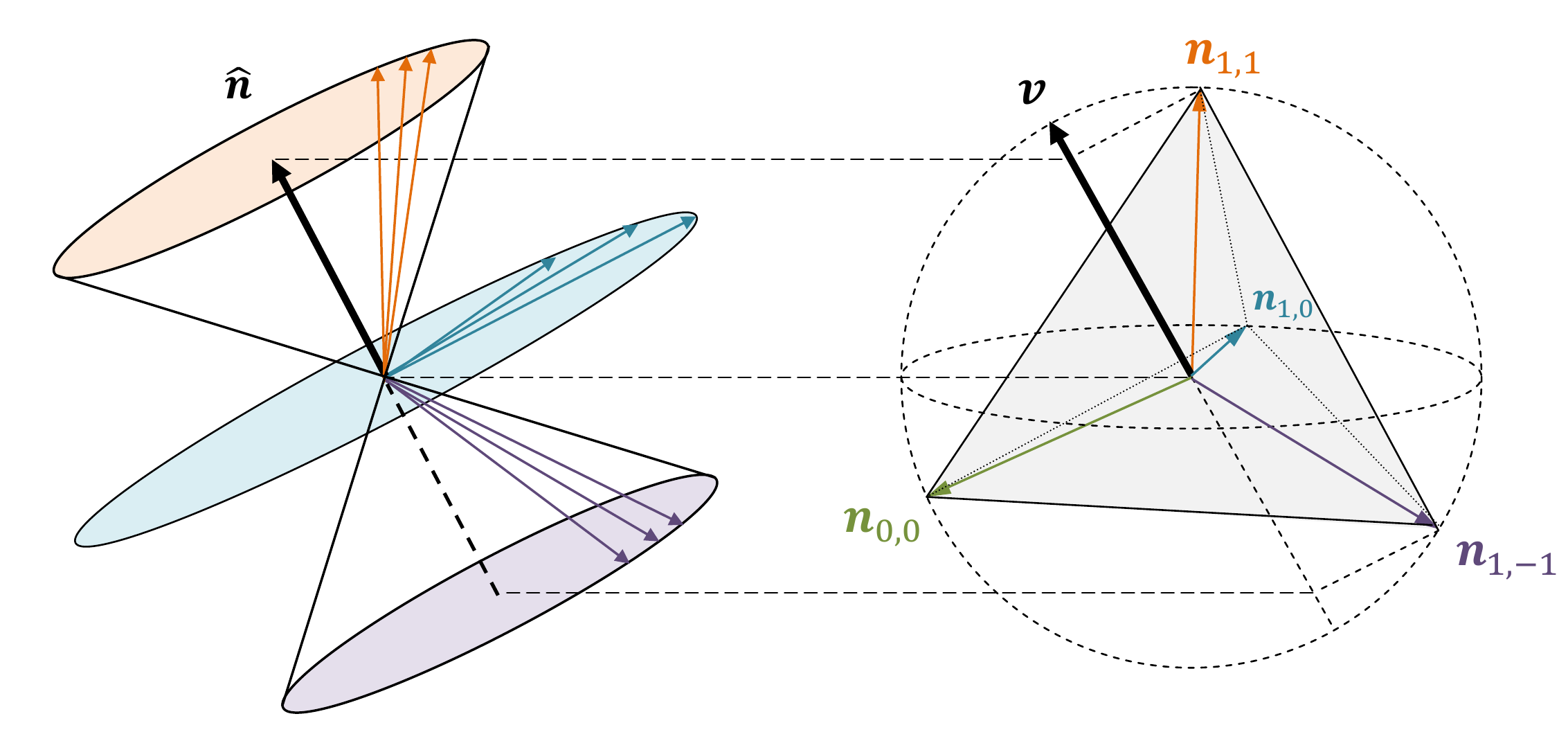}
\caption{The correspondence between the 3-dimensional classical theater and the 15-dimensional quantum theater, for an entity formed by the combination of two spin-${1\over 2}$ entities. Note that both vectors ${\bf n}_{1,0}$ and ${\bf n}_{0,0}$ project onto ${\bf v}$ at the origin of the Bloch sphere (${\bf n}_{1,0}\cdot {\bf v}={\bf n}_{0,0}\cdot {\bf v}=0$). However, although ${\bf n}_{1,0}$ is represented in the classical theater by a disk of vectors of length $\hbar$, orthogonal to ${\hat{\bf n}}$, the vector ${\bf n}_{0,0}$ is only represented by the point at the origin. Note also that, contrary to what the figure may suggest, the sphere containing the measurement tetrahedron is not part of the 3-dimensional sub-sphere in which ${\bf v}$ moves, isomorphically to ${\hat{\bf n}}$. 
\label{correspondence2}}
\end{figure} 

\begin{figure}[!ht]
\centering
\includegraphics[scale =0.57]{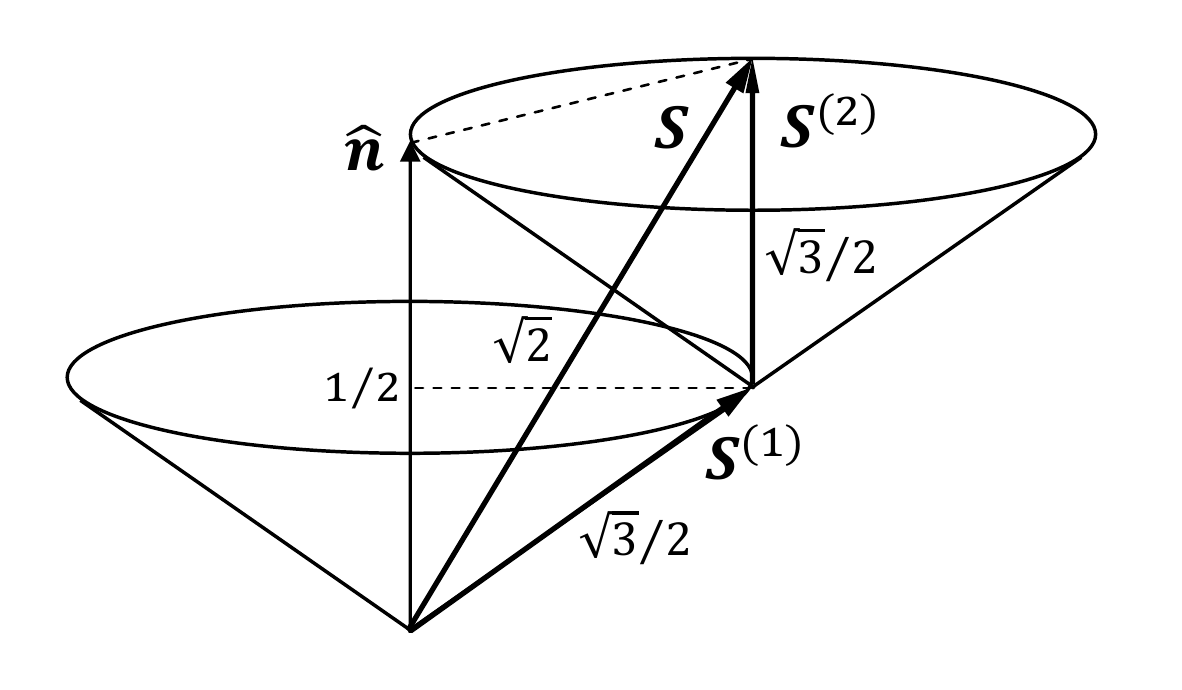}
\caption{The two Bloch eigenvectors ${\bf n}_{1,\pm1}$ can be represented in the classical theater by spin vectors ${\bf S}$ of length $\hbar \sqrt{2}$, such that ${\bf S}\cdot \hat{\bf n}=\pm1$ (in the drawing we have set $\hbar =1$, and only the ${\bf n}_{1,1}$ case is represented), which are the resultant of two suitably chosen vectors ${\bf S}^{(1)}$ and ${\bf S}^{(2)}$, both of length ${\sqrt{3}\over 2}$, and such that ${\bf S}^{(1)}\cdot \hat{\bf n}={\bf S}^{(2)}\cdot \hat{\bf n}=\pm{1\over 2}$. However, given an arbitrary vector ${\bf S}^{(1)}$ in the cone representative the first spin-${1\over 2}$ entity, not any vector ${\bf S}^{(2)}$ in the cone representative of the second spin-${1\over 2}$ entity will do the job. Indeed, some will produce a resultant vector ${\bf S}$ that is too short (the minimal length being $\hbar$), others too long (the maximal length being $\hbar\sqrt{3}$), and only two of them will be oriented exactly in such a way that ${\bf S}$ has the right length $\hbar \sqrt{2}$ (only one is represented in the figure).
\label{composingvectors}}
\end{figure}

When considering composite entities, the situation remains essentially the same, as we can still identify a vector ${\bf v}$ which is the referent, in the quantum theater, of the spatial direction $\hat{\bf n}$, as we have shown in \emph{Proposition 2}. In this case, however, the eigenstates of the composite entity cannot always be understood as resulting from combinations of the classical elements of reality associated with the composing entities. Considering the simple case of two spin-${1\over 2}$ entities (see Fig.~\ref{correspondence2}), we observe that the two Bloch vectors ${\bf n}_{1,\pm 1}$, associated with the \emph{product} (non-entangled) states $|\hat{\bf n},1,\pm1\rangle = |\hat{\bf n},\pm{1\over 2}\rangle\otimes |\hat{\bf n},\pm{1\over 2}\rangle$, can always be associated with classical angular momentum vectors of length $\hbar{\sqrt{3}\over 2}$ (compatibly with $(S^{(1)})^2\otimes\mathbb{I}^{(2)} |\hat{\bf n},1,\pm1\rangle = \mathbb{I}^{(1)}\otimes (S^{(2)})^2 |\hat{\bf n},1,\pm1\rangle = \hbar^2 {3\over 4} |\hat{\bf n},1,\pm1\rangle$), such that their projections along $\hat{\bf n}$ produce the values $\pm{\hbar\over 2}$ (compatibly with $S_{\hat{\bf n}}^{(1)}\otimes\mathbb{I}^{(2)}|\hat{\bf n},1,\pm1\rangle = \mathbb{I}^{(1)}\otimes S^{(2)}_{\hat{\bf n}}|\hat{\bf n},1,\pm1\rangle = \pm{\hbar\over 2} |\hat{\bf n},1,\pm1\rangle$), and their resultant is a vector of length $\hbar \sqrt{2}$ (compatibly with $S^2|\hat{\bf n},1,\pm1\rangle = \hbar^2 2 |\hat{\bf n},1,\pm1\rangle$) whose projection along $\hat{\bf n}$ gives $\pm\hbar$ (compatibly with $S_{\hat{\bf n}}|\hat{\bf n},1,\pm1\rangle = \pm\hbar |\hat{\bf n},1,\pm1\rangle$); see Fig.~\ref{composingvectors}.

On the other hand, the correspondence is not anymore complete for the Bloch vectors ${\bf n}_{1,0}$ and ${\bf n}_{0,0}$, which are representative of the \emph{entangled} (non-product) states $|\hat{\bf n},1,0\rangle = {1\over\sqrt{2}}(|\hat{\bf n},+{1\over 2}\rangle\otimes |\hat{\bf n},-{1\over 2}\rangle +|\hat{\bf n},-{1\over 2}\rangle\otimes |\hat{\bf n},+{1\over 2}\rangle)$ and $|\hat{\bf n},0,0\rangle = {1\over\sqrt{2}}(|\hat{\bf n},+{1\over 2}\rangle\otimes |\hat{\bf n},-{1\over 2}\rangle -|\hat{\bf n},-{1\over 2}\rangle\otimes |\hat{\bf n},+{1\over 2}\rangle)$, as is clear that they are not anymore eigenvectors of the one-entity spin observables $S^{(1)}_{\hat{\bf n}}\otimes\mathbb{I}^{(2)}$ and $\mathbb{I}^{(1)}\otimes S^{(2)}_{\hat{\bf n}}$. It is worth emphasizing, however, that the lack of a full correspondence between the quantum and classical theaters can also manifests when the composite system is in a state which is a product of eigenstates. For example, the two states $|\hat{\bf n},\pm{1\over 2}\rangle\otimes |\hat{\bf n},\mp{1\over 2}\rangle$, although they are eigenstates of the one-entity operators $S_{\hat{\bf n}}^{(1)}\otimes\mathbb{I}^{(2)}$, $(S^{(1)})^2\otimes\mathbb{I}^{(2)}$, $\mathbb{I}^{(1)}\otimes S_{\hat{\bf n}}^{(2)}$, $\mathbb{I}^{(1)}\otimes (S^{(2)})^2$, and of the total spin operator $S_{\hat{\bf n}}$, they are not however eigenstates of $S^2$, and therefore cannot be represented as classical three-dimensional vectors, as this would imply a well defined value for  $S^2$.

This difficulty, of obtaining a fully consistent correspondence for all spin eigenstates, is related to the well-known fact that we can either choose $\{(S^{(1)})^{ 2}\otimes\mathbb{I}^{(2)}, \mathbb{I}^{(1)}\otimes (S^{(2)})^{ 2}, S^2, S_{\hat{\bf n}}\}$, as a \emph{complete set of commuting observables}, or $\{S_{\hat{\bf n}}^{(1)}\otimes\mathbb{I}^{(2)}, (S^{(1)})^2\otimes\mathbb{I}^{(2)}, \mathbb{I}^{(1)}\otimes S_{\hat{\bf n}}^{(2)}, \mathbb{I}^{(1)}\otimes (S^{(2)})^2, S_{\hat{\bf n}}\}$. In the first case, we have the basis of spin eigenvectors $\{|\hat{\bf n},1,1\rangle, |\hat{\bf n},1,-1\rangle, |\hat{\bf n},1,0\rangle, |\hat{\bf n},0,0\rangle \}$, and we can always exhibit a correspondence for the total spin, but not always a correspondence for its components. In the second case, we  have the basis of spin eigenvectors $\{ |\hat{\bf n},+{1\over 2}\rangle\otimes |\hat{\bf n},+{1\over 2}\rangle, |\hat{\bf n},-{1\over 2}\rangle\otimes |\hat{\bf n},-{1\over 2}\rangle, |\hat{\bf n},+{1\over 2}\rangle\otimes |\hat{\bf n},-{1\over 2}\rangle, |\hat{\bf n},-{1\over 2}\rangle\otimes |\hat{\bf n},+{1\over 2}\rangle\}$, and we can always exhibit a correspondence for the composing spins, but not always a correspondence for the total spin. Note that the above two eigenbases share the first two of their vectors, which are precisely those eigenstates that admit a full representation, within the spatial theater, for both the total spin and for its components. 

According to the above, it is clear that an alternative, equivalent version of \emph{Proposition 2} can be written, using a different set of spin eigenvectors to defne the space vectors ${\bf v}$. More precisely, we have: 
\\

\noindent {\bf Proposition 2-bis}. \emph{Given two spin entities, of spin $s_1={N_1-1\over 2}$ and $s_2={N_2-1\over 2}$, and the spin observables $S_{\hat{\bf n}}^{(1)}$ and $S_{\hat{\bf n}}^{(2)}$, oriented along the space direction $\hat{\bf n}$, acting on $\compl^{N_1}$ and $\compl^{N_2}$, respectively, and given the total spin observable $S_{\hat{\bf n}}= S_{\hat{\bf n}}^{(1)}\otimes \mathbb{I}^{(2)} + \mathbb{I}^{(1)}\otimes S_{\hat{\bf n}}^{(2)}$, acting on $\compl^{N_1}\otimes \compl^{N_2}\equiv \compl^{N_1N_2}$, then, to the $N=N_1N_2 = (2s_1+1)(2s_2+1)$ eigenstates $|\hat{\bf n},\mu_1,\mu_2\rangle  =|\hat{\bf n},\mu_1\rangle\otimes |\hat{\bf n},\mu_2\rangle\in \compl^{N}$, such that 
$S_{\hat{\bf n}}^{(1)}\otimes \mathbb{I}^{(2)}|\hat{\bf n},\mu_1,\mu_2\rangle =\hbar\mu_1 |\hat{\bf n},\mu_1,\mu_2\rangle$, $\mathbb{I}^{(1)}\otimes S_{\hat{\bf n}}^{(2)}|\hat{\bf n},\mu_1,\mu_2\rangle =\hbar\mu_2 |\hat{\bf n},\mu_1,\mu_2\rangle$, and 
$S_{\hat{\bf n}}|\hat{\bf n},\mu_1,\mu_2\rangle =\hbar(\mu_1+\mu_2) |\hat{\bf n},\mu_1,\mu_2\rangle$, $\mu_1 =-s_1,\dots,s_1$, $\mu_2 =-s_2,\dots,s_2$, represented in the extended Bloch sphere by the unit vectors ${\bf n}_{\mu_1,\mu_2}\in B_1(\real^{N^2-1})$, we can associate a unit vector ${\bf w}\in B_1(\real^{N^2-1})$, defined by:
\begin{equation}
{\bf w} =d_{N_1,N_2} \sum_{\mu_1=-s_1}^{s_1}\sum_{\mu _2=-s_2}^{s_2}(\mu _1+\mu_2)\, {\bf n}_{\mu_1,\mu_2},\quad d_{N_1,N_2}=\sqrt{12(N-1)\over NN_1N_2 (N_1^2 + N_2^2-2)},
\label{vectorv2-tris}
\end{equation}
such that when the three-dimensional unit vector $\hat{\bf n}$ runs through all the possible directions of the Euclidean space, ${\bf w}$ equally spans, in an isomorphic way, all the points at the surface of a $3$-dimensional sub-ball of $B_1(\real^{N^2-1})$.
}\\

The proof of  \emph{Proposition 2-bis} is identical -- \emph{mutatis mutandis} -- to that of \emph{Poroposition 2}. One may wonder if the vectors ${\bf v}$ defined in (\ref{vectorv2-bis}) are the same as the vectors ${\bf w}$ defined in (\ref{vectorv2-tris}), considering that they both obey (\ref{scalarproduct}). The answer is affirmative, and to show this it is sufficient to observe that:
\begin{eqnarray}
{c_N\over N d_{N_1,N_2}}\,{\bf w}\cdot\mbox{\boldmath$\Lambda$} &=& \sum_{\mu_1=-s_1}^{s_1}\sum_{\mu _2=-s_2}^{s_2}(\mu _1+\mu_2)\, {c_N\over N}{\bf n}_{\mu_1,\mu_2}\cdot\mbox{\boldmath$\Lambda$}
=\sum_{\mu_1=-s_1}^{s_1}\sum_{\mu _2=-s_2}^{s_2}(\mu _1+\mu_2)\, {1\over N}(\mathbb{I} +c_N{\bf n}_{\mu_1,\mu_2}\cdot\mbox{\boldmath$\Lambda$})\nonumber\\
&=& {1\over\hbar}\sum_{\mu_1=-s_1}^{s_1}\sum_{\mu _2=-s_2}^{s_2}\hbar (\mu _1+\mu_2) P({\bf n}_{\mu_1,\mu_2})={1\over\hbar}S_{\hat{\bf n}},
\end{eqnarray}
where for the second equality we have used that $\sum_{\mu_1=-s_1}^{s_1}\sum_{\mu _2=-s_2}^{s_2}(\mu _1+\mu_2)=0$. In other terms, modulo a renormalization factor, the components of ${\bf w}$ provide the expansion of $S_{\hat{\bf n}}$, on the basis of the generators of $SU(N)$. But since we also have: 
\begin{eqnarray}
{c_N\over N d_{N_1,N_2}}\,{\bf v}\cdot\mbox{\boldmath$\Lambda$} &=& \sum_{s = |s_1-s_2|}^{s_1+s_2}\sum_{\mu _s=-s}^{s}\mu _s\, {c_N\over N}{\bf n}_{s,\mu_s}\cdot\mbox{\boldmath$\Lambda$}
= \sum_{s = |s_1-s_2|}^{s_1+s_2}\sum_{\mu _s=-s}^{s}\mu _s\, {1\over N}(\mathbb{I} +c_N{\bf n}_{s,\mu_s}\cdot\mbox{\boldmath$\Lambda$})\nonumber\\
&=& {1\over\hbar}\sum_{s = |s_1-s_2|}^{s_1+s_2}\sum_{\mu _s=-s}^{s}\hbar \mu _s P({\bf n}_{s,\mu_s})={1\over\hbar}S_{\hat{\bf n}},
\end{eqnarray}
and the expansion of $S_{\hat{\bf n}}$ is unique, we can conclude that the vectors (\ref{vectorv2-bis}) and (\ref{vectorv2-tris}) are exactly the same.

\section{Concluding remarks}
\label{Conclusion}

In this article we have presented the extended Bloch representation of quantum mechanics and the associated hidden-measurement interpretation, and used it to explore spin states and spin measurements. The extended Bloch representation is a candidate for a refined version of quantum theory, in which operator-states (density matrices) can also play a role as pure states in the description of what goes on during a measurement process. We refer the reader to \cite{AertsSassoli2014c} for a more detailed and complete description of the model, and for a further generalization of it. 

Let us mention that it is also possible to relax the hypothesis that the substance filling the measurement simplexes is uniform (in the sense of having a uniform probability density of disintegrating in a point), and still be able to derive the Born rule, if a (universal) average over all possible non-uniform substances is considered. Also, one can show that non-uniform substances, when not averaged out, describe physical entities of a genuine intermediate nature, which cannot be described by the classical or quantum formalisms (their elements of reality being neither contained in the classical nor in the quantum theaters)~\cite{AertsSassoli2014c,AertsSassoli2014a,AertsSassoli2014b}. 

The extended Bloch representation elucidates the mechanism through which quantum entities interact with classical measurement apparatus. The interaction is invasive and governed by (non-spatial) fluctuations, which by definition cannot be controlled by the experimenter, and therefore produce a genuine indeterministic change of the state of the measured entity. According to this interpretation, a quantum measurement is not only a process of discovery, but also, and above all, a process of creation. The extended Bloch representation also allows us to gain some insight into how the vaster quantum world relates to our ordinary ``niche'' reality, made of macroscopic objects which are always present in the three-dimensional Euclidean theater. In that respect, we have shown that quantum spins cannot be considered as intrinsic angular momenta, as no space directions can be directly associated to them, beyond the $s={1\over 2}$ situation. On the other hand, they always remain in a specific relation to the space directions, as expressed by the fact that they possess a predetermined orientation with respect to the vectors (\ref{vectorv}) and (\ref{vectorv2-bis})-(\ref{vectorv2-tris}), which are the representative of the space directions within the Blochean theater.

It is worth observing that even though spin-${1\over 2}$ entities can be characterized, when isolated, by single space directions, not for this they should be considered to be generally in space, as is clear that $SU(2)$ is a \emph{double cover} of $SO(3)$, which means that $2\pi$ and $4\pi$ rotations, although they correspond to two distinct elements of $SU(2)$, are mapped onto the same element of $SO(3)$. In the Bloch sphere this two-to-one correspondence cannot be seen, as global phase factors are not represented into it. However, when a spin is coupled with the translational degrees of freedom of a microscopic entity, the distinction between $2\pi$ and $4\pi$ rotations can become observable, in the suitable experimental context.

\begin{figure}[!ht]
\centering
\includegraphics[scale =0.57]{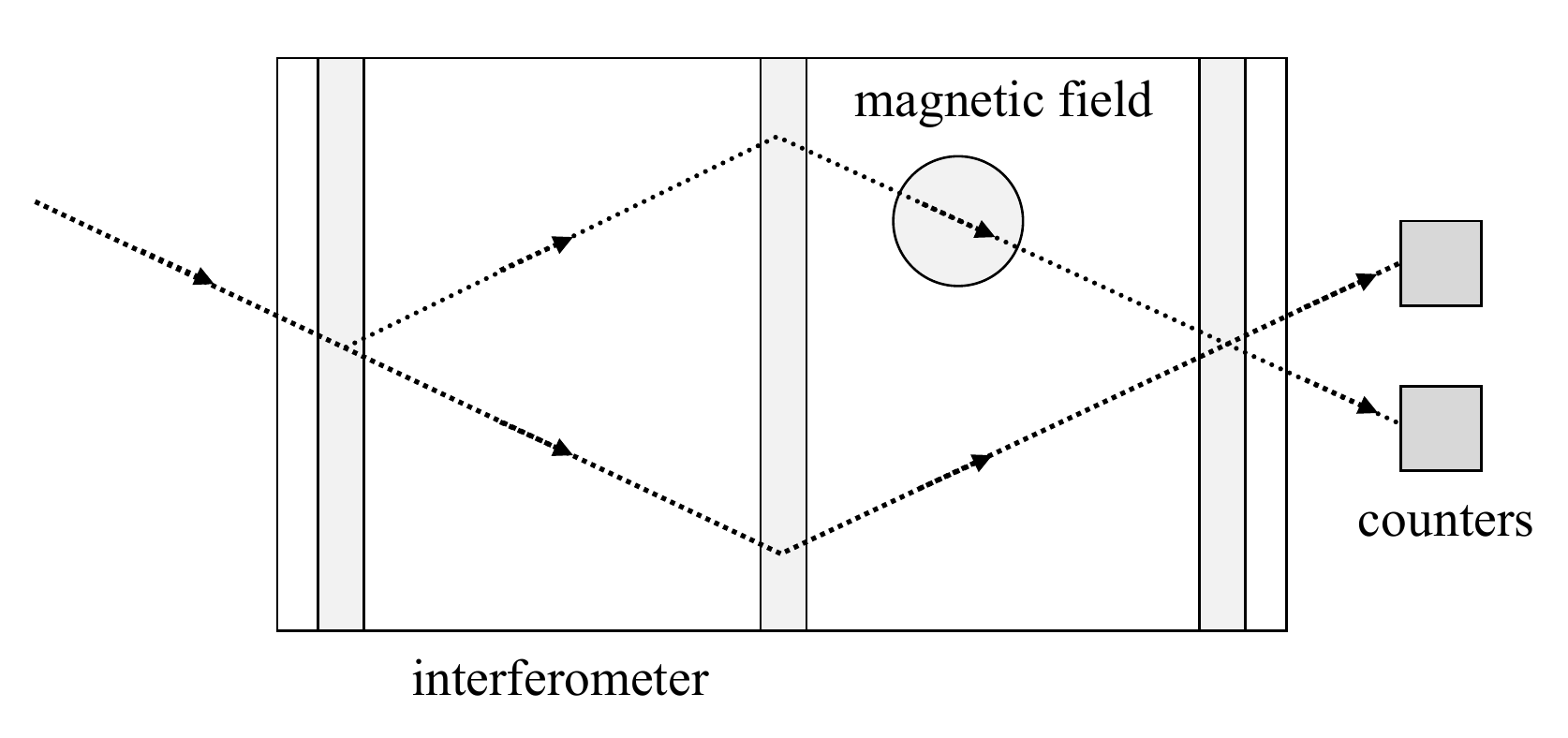}
\caption{Sketch of a neutron interferometry experiment, showing the two paths of the coherently split wave function, and the magnetic field of variable intensity which is locally applied on one of them.
\label{Rauch-figure}}
\end{figure}

This is precisely what was achieved by Helmut Rauch and his group, in their celebrated experiments of interferometry~\cite{Rauch1975}, where neutrons, passing one by one through a perfect silicon crystal (made of three parallel lips of same thickness; see Fig.~\ref{Rauch-figure}), were allowed to interfere with themselves, because of the spatial splitting and subsequent recombination of their wave functions. This was done in a way so as to allow the experimenter to only act on one of the two spatially separated components, by means of a magnetic field whose intensity could be varied, so producing an interference contribution in the detected intensity, with a remarkable $4\pi$-periodic behavior (when expressed as a function of the rotation angle of the neutron's spin). 

Rauch's experiment has shown us two things: the first one is that, as we said, even spin-${1\over 2}$ entities are not in space, considering that a $2\pi$ rotation is not sufficient to bring them back exactly in the same state, when not in an isolated condition. In fact, this was already clear from our discussion in the Introduction, considering that we could not find classical elements of reality able to account for spin observations along all possible directions. And this also emerged in our analysis of the combination of two spin--${1\over 2}$ entities, which although they can be in a well defined eigenstate within the composite entity (when in a product state), not for this they can always be associated with a well-defined three-dimensional classical angular momentum vectors. 

Of course, if, strictly speaking, spins are not in space, it means that it is not totally correct to interpret the action of the magnetic field on the spinor component of the wave function in Rauch's experiments as a true rotation in space. And in fact, it has been proposed (also by one of us) that a minimal interpretation of the experiment is as a longitudinal Stern-Gerlach effect, i.e., that the (apparent) Larmor precession would only result from the interference between the two Stern-Gerlach states in the weak field limit~\cite{Mezei1988, Martin1994}. 

Rauch's experiment has also shown us that the notion of non-spatiality equally applies to the configuration space part of the wave function. Indeed, it shows that although a neutron can only be detected in one place, it nevertheless can be acted upon, simultaneously, from different separated places, in a way that can produce a measurable effect. And if we reflect attentively, such a possibility can only be understood if we admit that a neutron is an entity which is not permanently present in space, but only possibly enters it when it interacts with a measuring apparatus, which in a sense is able to ``drag'' it into a spatial state~\cite{Aerts1999}. 

From a conceptual point of view, we should certainly distinguish the notion of \emph{space direction} from the notion of \emph{space location}. Elementary entities are generally to be considered not in space in both senses: because they don't generally have an actual position in space, but only a potential one, and because their spin (if they are not zero) are not oriented toward a space direction, apart the special case of isolated spin-${1\over 2}$ entities (where the term ``isolated'' means here not coupled with other systems, including the translational degrees of freedom of the very entity carrying the spin). Now, as we said, microscopic entities are able to acquire a location in space, in specific contexts, as is clear that we can localize them in space with arbitrary precision. On the other hand, their spins, apart the $s={1\over 2}$ case, remain always out of alignment with respect to space, as is clear that all spin eigenvectors within the Bloch sphere are oriented in directions which are different from those of the vectors ${\bf v}$, defined in \emph{Proposition 1, 2} and \emph{2-bis}, which are the referents of the space directions within the quantum theater. 

In the special $s={1\over 2}$ situation, the space directions and the directions of the spin eigenstates coincide, but as we have seen the correspondence between spin eigenvectors and angular momentum vectors is incomplete: because it can only be established for a single spin measurement at a time, and because it doesn't fully work when composite entities are considered. Therefore, even spin-${1\over 2}$ should not be considered to be in space (or fully in space), i.e., to be representable by one (or many) three-dimensional vectors of suitable length and orientation. 

But although generally not in space, quantum spin entities, when in an eigenstate, are nevertheless always in a specific relation to space, as evidenced by the fact that the eigenvectors in the Bloch sphere always maintain fixed and predetermined orientations with respect to the ``spatial'' vectors ${\bf v}$. These orientations are precisely those that allow the eigenvectors to always be positioned at the vertices of the measurement simplexes, so that they cannot  be affected by the collapse of the corresponding membranes. This is why we can predict in advance the outcome of the measurements, and bring them into correspondence with the classical angular momentum elements of reality of the classical theater. It is in that sense, and only in that sense, that we can justify the assertion that, say, an electron has a spin of a given magnitude along a given direction. 

However, as emphasized many times, when we combine spins, the classical picture breaks down, as is clear that there is no simple relation between the elements of reality described by two 3-dimensional Bloch spheres, associated with two separated spin-${1\over 2}$ entities, and the 15-dimensional Bloch sphere describing the states emerging from their quantum combinations. This because no sum of three-dimensional vectors will ever be able to account  for the emergence of a 15-dimensional Bloch sphere from two 3-dimensional ones, and of a tetrahedron (3-simplex) from two line segments (1-simplexes). 

A final comment is in order. As we explained in Sec.~\ref{Connecting elements of reality}, not all the elements of reality that are present in the quantum theater do necessarily have a correspondence, however partial, with some of the elements of reality present in the classical theater. Quantum spins can individually be observed in the classical theater also because they are all in a specific relation to space directions, i.e., to classical elements of reality. But this needs not to be the case for other quantum properties, like for instance the color charge properties of individual quarks, which, as far as we know, and contrary to spins, have no relations to space directions, which may be one of the reasons of their confinement within the quantum cave. 

This is in accordance with the view of realism that we have adopted in this article, that we will call \emph{multiplex realism}. It is worth emphasizing that multiplex realism is not instrumentalism in disguise, nor a new form of model-dependent realism~\cite{Hawking2010}, or of  structural realism~\cite{Worrall1989}. Also, the view should not be considered as the expression of the belief that we necessarily live in a multiplex reality. This is certainly a possibility, but it should be clear that the main reason for adopting such a concept of realism is that we give importance to the following three observations.

(1) Due to our presence, for hundreds of thousands of years, in that specific niche of reality which is the surface of our planet Earth, surrounded by material objects that with good approximation obey the Newton's laws, we have constructed an Euclidean theater to stage our relations and interactions with these classical entities. 

(2) In more recent times, we became aware of the existence of quantum and relativistic entities, whose reality could only be put in a partial correspondence with the properties and behaviors of the previously known Euclidean entities, that is, we could only find \emph{partial morphisms}, and not perfect isomorphisms. For instance, quantum eigenstates could be linked, to some extent, to classical elements of reality, but not their superpositions. 

(3) Although we have good reasons to believe that the quantum and relativistic theaters are the expression of more advanced theories of reality, we nonetheless continue to be strongly influenced by our Euclidean theater, also because we are ordinarily not aware that it has been constructed by our ancestors, in particular as regards the primitive notions of space and time. Therefore, we have interest in becoming fully aware of this `construction aspect,' taking into consideration its effect in the way we conceive, and try to understand, the more recently discovered non-Euclidian quantum and relativistic elements of reality.

However, it is important to emphasize that realism can be multiplex even if reality is singleplex. By this we mean that multiplex realism should not to be intended as a statement about the multiple nature of our reality, but as an approach to reality which considers that certain incompatibilities, like that between relativity and quantum theories, can be more fruitfully studied if we  take into account the fact that their elements of reality are also, in part, the result of a constructive process.


\begin{thebibliography}{}
\setlength{\itemsep}{-0.8mm}
\bibitem{Aerts2002} D. Aerts, ``The unification of personal presents: a dialogue of different world views,'' pp. 63--82. In: \emph{International Readings on Theory, History and Philosophy of Culture: Ontology of Dialogue}, Vol.12 (2002).
\bibitem{Stern1922} W. Gerlach and O. Stern, ``Das magnetische Moment des Silberatoms,'' Zeitschrift für Physik 9, 353--355 (1922).
\bibitem{Rauch1974} H. Rauch, W. Treimer, U. Bonse, ``Test of a single crystal neutron interferometer,'' Phys. Lett. A 47, 369 (1974).
\bibitem{Rauch1975} H. Rauch et al., ``Verification of coherent spinor rotations of fermions," Phys. Lett. 54A, 425--427 (1975).
\bibitem{Rauch1988} H. Rauch, ``Neutron interferometric tests of quantum mechanics," Helv. Phys. Acta 61, 589 (1988).
\bibitem{Aspect1982a} A. Aspect et al., ``Experimental Realization of Einstein-Podolsky-Rosen-Bohm Gedankenexperiment: A New Violation of Bell's Inequalities,'' Phys. Rev. Lett., 49, p. 91 (1982).
\bibitem{Aspect1999} A. Aspect, ``Bell's inequality test: more ideal than ever,'' Nature (London), 398, p. 189 (1999). 
\bibitem{Hentschel2009} K. Hentschel, ``Spin,'' pp. 726--731. In: \emph{Compendium of Quantum Physics}, D. Greenberger, K. Hentschel and F. Weinert (eds.), Springer-Verlag, Berlin Heidelberg (2009).
\bibitem{Poincare1892} H. Poincar\'e, \emph{Th\'eorie Mathematique de la Lumi\`ere}, Gauthiers-Villars, Paris, Vol. 2 (1892).
\bibitem{Bloch1946} F. Bloch, ``Nuclear induction,'' Phys. Rev. 70, 460--474 (1946).
\bibitem{Aerts1986} D. Aerts, ``A possible explanation for the probabilities of quantum mechanics,'' Journal of Mathematical Physics 27, 202--210 (1986).
\bibitem{Aerts1987} D. Aerts, ``The origin of the non-classical character of the quantum probability model.'' In: Information, Complexity, and Control in Quantum Physics, eds. A. Blanquiere et al, Springer-Verlag, Berlin (1987).
\bibitem{AertsSassoli2014c} D. Aerts and M. Sassoli de Bianchi, ``The Extended Bloch Representation of Quantum Mechanics and the Hidden-Measurement Solution to the Measurement Problem,'' Annals of Physics 351, 975--1025 (2014). DOI: 10.1016/j.aop.2014.09.020.
\bibitem{Aerts1998b} D. Aerts, ``The entity and modern physics: the creation-discovery view of reality.'' In: \emph{Interpreting Bodies: Classical and Quantum Objects in Modern Physics}, ed. Castellani, E. Princeton Unversity Press, Princeton (1998).
\bibitem{Aerts1999} D. Aerts, ``The Stuff the World is Made of: Physics and Reality,'' pp. 129--183. In: \emph{The White Book of `Einstein Meets Magritte'}, Edited by Diederik Aerts, Jan Broekaert and Ernest Mathijs, Kluwer Academic Publishers, Dordrecht, 274 pp. (1999).
\bibitem{Gleason1957} A. M. Gleason, ``Measures on the closed subspaces of a Hilbert space,'' J. Math. Mech. 6, 885--893 (1957). 
\bibitem{Kochen1967} S. Kochen and E. P. Specker, ``The problem of hidden variables in quantum mechanics,'' J. Math. Mech. 17, 59--87 (1967).
\bibitem{Aertsetal1997} D. Aerts, B. Coecke, B. D. Hooghe and F. Valckenborgh, ``A mechanistic macroscopic physical entity with a three-dimensional Hilbert space description,'' Helv. Phys. Acta 70, 793 (1997).
\bibitem{Coecke1995} B. Coecke, ``Hidden measurement representation for quantum entities described by finite dimensional complex Hilbert spaces,'' Found. Phys., 25, 1185 (1995).
\bibitem{Coecke1995b} B. Coecke, ``Generalization of the proof on the existence of hidden measurements to experiments with an infinite set of outcomes,'' Found. Phys. Lett., 8, 437 (1995).
\bibitem{Arvind1997} Arvind, K. S. Mallesh and N. Mukunda, ``A generalized Pancharatnam geometric phase formula for three-level quantum systems,'' J. Phys. A 30, 2417 (1997)
\bibitem{Kimura2003} G. Kimura``The Bloch Vector for $N$-Level Systems,'' Phys. Lett. A 314, 339 (2003).
\bibitem{Byrd2003} M. S. Byrd and N. Khaneja, ``Characterization of the positivity of the density matrix in terms of the coherence vector representation,'' Phys. Rev. A 68, 062322 (2003).
\bibitem{Kimura2005} G. Kimura \& A. Kossakowski, ``The Bloch-vector space for N-level systems -- the spherical-coordinate point of view,'' Open Sys. Information Dyn. 12, 207 (2005).
\bibitem{Bengtsson2006} I. Bengtsson and K. \.{Z}yczkowski, \emph{Geometry of quantum states: An introduction to quantum entanglement}, Cambridge University Press, Cambridge (2006).
\bibitem{Bengtsson2013} I. Bengtsson and K. \.{Z}yczkowski, ``Geometry of the Set of Mixed Quantum States: An Apophatic Approach,'' pp. 175--197. In: \emph{Geometric Methods in Physics, XXX Workshop 2011, Trends in Mathematics}, Springer (2013).
\bibitem{Hughston1993} L. P. Hughston, R. Jozsa and William K. Wootters,``A complete classification of quantum ensembles having a given density matrix,'' Physics Letters A 183, 14--18 (1993).
\bibitem{Hioe1981} F. T. Hioe, J. H. Eberly, ``$N$-Level Coherence Vector and Higher Conservation Laws in Quantum Optics and Quantum Mechanics,'' Phys. Rev. Lett. 47, 838--841 (1981).
\bibitem{Alicki1987} R. Alicki, K. Lendi, \emph{Quantum Dynamical Semigroups and Application}, Lecture Notes in Physics Vol. 286, Springer-Verlag, Berlin (1987).
\bibitem{Mahler1995} G. Mahler, V. A. Weberruss, \emph{Quantum Networks}, Springer, Berlin (1995).
\bibitem{Aerts1982b} D. Aerts ``Description of many physical entities without the paradoxes encountered in quantum mechanics,'' Found. Phys. 12, 1131--1170 (1982).
\bibitem{Aerts1984} D. Aerts, ``The missing elements of reality in the description of quantum mechanics of the EPR paradox situation,'' Helvetica Physica Acta, 57, 421--428 (1984).
\bibitem{Aerts1999b} D. Aerts, D. ``Foundations of quantum physics: a general realistic and operational approach,'' International Journal of Theoretical Physics 38, 289--358 (1999).
\bibitem{Aerts2000} D. Aerts, ``The description of joint quantum entities and the formulation of a paradox,'' International Journal of Theoretical Physics 39, 485--496 (2000).
\bibitem{Einstein1935} A. Einstein, B. Podolsky and N. Rosen, ``Can Quantum-Mechanical Description of Physical Reality Be Considered Complete?,'' Phys. Rev. 47, 777 (1935).
\bibitem{AertsSassoli2014a} D. Aerts and M. Sassoli de Bianchi, ``The unreasonable success of quantum probability I: Quantum measurements as uniform measurements,'' arXiv:1401.2647 [quant-ph] (2014).
\bibitem{AertsSassoli2014b} D. Aerts and M. Sassoli de Bianchi, ``The unreasonable success of quantum probability II: Quantum measurements as universal measurements,'' arXiv:1401.2650 [quant-ph] (2014).
\bibitem{Aerts1990} D. Aerts, ``An attempt to imagine parts of the reality of the micro-world,'' 3--25, in \emph{Problems in Quantum Physics II}; Gdansk '89, Edited by: J. Mizerski et al., World Scientific Publishing Company, Singapore (1990).
\bibitem{Aerts1991} D. Aerts, ``A mechanistic classical laboratory situation violating the Bell inequalities with $2\sqrt{2}$, exactly `in the same way' as its violations by the EPR experiments,'' Helv. Phys. Acta 64, 1--23 (1991).
\bibitem{Sassoli2013b} M. Sassoli de Bianchi, ``Using simple elastic bands to explain quantum mechanics: a conceptual review of two of Aerts' machine-models,'' Centr. Eur. J. Phys. 11, 147--161 (2013).
\bibitem{Mezei1988} F. Mezei, ``Zeeman energy, interference and neutron spin echo: a minimal theory,'' Physica B 151, 74--81 (1988). 
\bibitem{Martin1994} Ph. A. Martin and M. Sassoli de Bianchi, ``Spin precession revisited,'' Found. Phys. 24, 1371--1378 (1994).
\bibitem{Hawking2010} S. Hawking and L. Mlodinow, \emph{The Grand Design}, Bantam Books, New York (2010).
\bibitem{Worrall1989} J. Worrall, J., 1989. ``Structural realism: The best of both worlds?'' Dialectica 43, 99--124. Reprinted in D. Papineau (ed.), \emph{The Philosophy of Science}, Oxford University Press, Oxford, pp. 139--165 (1989).





\end{thebibliography}
\end{document}